\documentclass[a4paper]{cas-dc}
\usepackage[numbers,sort&compress]{natbib}
\usepackage{braket}
\usepackage{multicol}
\usepackage{subfigure}

\usepackage{mathcomp} 
\usepackage[normalem]{ulem}
\usepackage{cancel}

\definecolor{darkpastelgreen}{rgb}{0.01, 0.75, 0.24}
\definecolor{electricindigo}{rgb}{0.44, 0.0, 1.0}
\definecolor{palatinateblue}{rgb}{0.15, 0.23, 0.89}
\definecolor{carminered}{rgb}{1.0, 0.0, 0.22}
\definecolor{markbrown}{rgb}{0.45, 0.26, 0.08}
\definecolor{markgray}{gray}{0.55}
\hypersetup{linkcolor=palatinateblue, citecolor=electricindigo, urlcolor=carminered}

\def\be{\begin{eqnarray}}\def\ee{\end{eqnarray}}

\begin{document}
\shorttitle{
Gravitational $D$-form factors of the nucleon and pion with mixing between two-quark and four-quark states
}
\shortauthors{M. Kawaguchi, Y. Ma and Y. L. Ma}  
\title[mode = title]{
Gravitational $D$-form factors of the nucleon and pion with mixing between two-quark and four-quark states
}  

\author[1]{Mamiya Kawaguchi}
\nonumnote{$\dag$ E-mail: 
mamiya@aust.edu.cn
(corresponding author)
}

\author[2]{Yao Ma}
\nonumnote{$\ddag$ E-mail: 
mayao@nju.edu.cn
}

\author[3,2]{Yong-Liang Ma}
\nonumnote{$\S$ E-mail: 
ylma@nju.edu.cn
}

\address[1]{
Center for Fundamental Physics, School of Mechanics and Photoelectric Physics,
Anhui University of Science and Technology, Huainan, 232001, China
}
\address[2]{School of Frontier Sciences, Nanjing University, Suzhou 215163, China}
\address[3]{School of Physics, Nanjing University, Nanjing 210093, China}

\begin{abstract}
We investigate the gravitational $D$-form factors of the nucleon and pion within an extended linear sigma model (eLSM) that incorporates both two-quark and four-quark states, as well as nucleon fields. 
Within the scalar-meson-dominance picture, we find that meson mixing between two-quark and four-quark states suppresses the nucleon $D$-form factor and leads to agreement with the lattice-QCD results. Interestingly, a quantitatively similar suppression was previously found in the parity doublet model (PDM) as a consequence of a nonzero chiral invariant component of the nucleon mass. This suggests that the effects of meson mixing and the chiral invariant mass are difficult to distinguish from the nucleon $D$-form factor alone.
We therefore examine the pion $D$-form factor in the chiral limit. 
While its value in the forward limit remains fixed at $-1$ in both the eLSM and the PDM, meson mixing modifies its momentum-transfer dependence.
In contrast, the chiral invariant mass does not directly affect the pion $D$-form factor.
These results suggest that a combined analysis of the nucleon and pion $D$-form factors can provide further insight into the mass decomposition of the nucleon and the quark composition of mesons.

\end{abstract}


\maketitle

\section{Introduction}

Understanding the origin of hadron masses and elucidating how hadrons are composed in terms of quark degrees of freedom are longstanding fundamental questions in the low-energy regime of QCD. Gravitational form factors (GFFs) of hadrons provide valuable information on their internal structure~\cite{Polyakov:2018zvc,Burkert:2023wzr} and may therefore serve as useful probes for addressing these questions. In recent years, GFFs have attracted increasing attentions and have been investigated through various models~\cite{Cebulla:2007ei,Hudson:2017xug,Kumano:2017lhr,Polyakov:2018exb,Lorce:2018egm,Hatta:2018sqd,Tanaka:2018wea,Anikin:2019kwi,Avelino:2019esh,Yanagihara:2019foh,Hatta:2019lxo,Freese:2019eww,Azizi:2019ytx,Mamo:2019mka,Neubelt:2019sou,Krutov:2020ewr,Alharazin:2020yjv,Varma:2020crx,Kim:2020nug,Chakrabarti:2020kdc,Yanagihara:2020tvs,Kim:2020lrs,Tong:2021ctu,Tong:2022zax,Freese:2021czn,Panteleeva:2021iip,Hatta:2021can,Mamo:2021krl,Freese:2021qtb,Gegelia:2021wnj,Kim:2021jjf,Owa:2021hnj,Lorce:2021xku,Ji:2021mfb,More:2021stk,Mamo:2022eui,Lorce:2022cle,Fujita:2022jus,Choudhary:2022den,Kim:2022wkc,Alharazin:2022wjj,Won:2022cyy,Tanaka:2022wzy,Xu:2023izo,Ito:2023oby,Lorce:2023zzg,Amor-Quiroz:2023rke,Guo:2023pqw,Won:2023ial,Guo:2023qgu,Czarnecki:2023yqd,Won:2023zmf,Hatta:2023fqc,Liu:2023cse,Liu:2024jno,Broniowski:2024oyk,Cao:2024zlf,Liu:2024rdm,Yao:2024ixu,Fujii:2024rqd,Goharipour:2025lep,Dehghan:2025ncw,Goharipour:2025yxm,Broniowski:2025ctl,Hatta:2025vhs,Hatta:2025ryj,Dehghan:2025eov,Guo:2025jiz,Liu:2025vfe,Sugimoto:2025btn,Nair:2025sfr,Cao:2025dkv,Coriano:2024wrz,Coriano:2025lge,Stegeman:2025sca,Fujii:2025aip,Tanaka:2025pny,Fukushima:2025jah,Tanaka:2025hhv,Stegeman:2025tdl,Xing:2025uwn,Kawaguchi:2025cuf,Fukushima:2026wwc,Ejima:2026bzo,Deng:2026ydm,Hagiwara:2024wqz,Kawaguchi:2026yae}, lattice QCD simulations~\cite{Shanahan:2018pib,Shanahan:2018nnv,Pefkou:2021fni,Hackett:2023nkr,Hackett:2023rif,Wang:2024lrm}, and experiments~\cite{Burkert:2018bqq,Burkert:2021ith,Duran:2022xag,CLAS:2026lls,Dutrieux:2021nlz}.

Focusing on the nucleon, its gravitational form factors (GFFs) are defined through the matrix element of the energy-momentum tensor (EMT),
\begin{equation}
\begin{split}
&\langle N(p',s')|\Theta^{\mu\nu}(x)|N(p,s)\rangle
\\
&=
\bar u^{s'}(p')
\Bigg[
A_N(t)\frac{\bar P^\mu\bar P^\nu}{m_N}
+
J_N(t)
\frac{
i\sigma^{\mu\rho}q_\rho\bar P^\nu
+
i\sigma^{\nu\rho}q_\rho\bar P^\mu
}{2m_N}
\\
&\hspace{3cm}
+
D_N(t)
\frac{q^\mu q^\nu-g^{\mu\nu}q^2}{4m_N}
\Bigg]
u^s(p)e^{-iq\cdot x},
\end{split}
\end{equation}
where $\bar P^\mu=(p'^\mu+p^\mu)/2$, $q^\mu=p'^\mu-p^\mu$,
and $t=-q^2$.
The functions $A_N(t)$, $J_N(t)$ and $D_N(t)$
are the nucleon GFFs, which encode information on the mass, spin, and internal mechanical structure of nucleon. In particular, the $D$-form factor would be directly related to the internal forces and mechanical properties of the nucleon~\cite{Polyakov:2002yz,Polyakov:2018zvc,Lorce:2018egm,Lorce:2025oot}.
Despite its importance for understanding hadron structure, the $D$-form factor remains much less constrained than the other GFFs.
This leaves open the possibility that the $D$-form factor may reveal new aspects of hadron structure that are not readily accessible through other observables.

In a previous study by some of the present authors~\cite{Kawaguchi:2025cuf}, we investigated the nucleon D-form factor within the parity doublet model (PDM)~\cite{Detar:1988kn,Jido:2001nt}, in which the nucleon mass contains a chiral invariant component independent of spontaneous chiral symmetry breaking. We evaluated the nucleon $D$-form factor under the assumption of lightest-scalar-meson dominance. The scalar-meson-dominance picture has been considered in several previous studies and is expected to provide a reasonable description in the low-energy regime of QCD~\cite{Broniowski:2024oyk,Broniowski:2025ctl,Stegeman:2025sca,Stegeman:2025tdl,Kawaguchi:2025cuf}. Within this framework, we found that a nonzero chiral invariant mass suppresses the magnitude of the nucleon D-form factor, and comparison with lattice-QCD data~\cite{Hackett:2023rif} allowed us to constrain its value. Interestingly, the resulting range was found to be comparable to that constrained by neutron-star studies~\cite{Motohiro:2015taa,Minamikawa:2020jfj,Minamikawa:2023eky, Gao:2024lzu,Gao:2025nkg,Gao:2025vdc}, where the PDM has been used to describe neutron-star properties such as the mass-radius relation. This suggests that the nucleon $D$-form factor can serve as a useful probe for investigating the origin and detailed structure of the nucleon mass. However, it remains unclear whether such a suppression is uniquely associated with the chiral invariant mass or can also arise from other mechanisms. It is also worth noting that scalar tetraquark degrees of freedom have previously been incorporated into the PDM in studies of nuclear matter, where they were found to have non-negligible effects~\cite{Gallas:2011qp}. 
Since such degrees of freedom can modify the meson sector and thereby influence nuclear matter properties, it is natural to ask whether multiquark components may also affect the nucleon $D$-form factor.

In the present study, we investigate whether a similar suppression of the nucleon $D$-form factor can arise from the meson effect. For this purpose, we employ an extended linear sigma model (eLSM) containing both two-quark and four-quark states~\cite{Fariborz:2005gm,Fariborz:2007ai,Fariborz:2009cq,Ma:2023eoz,Ma:2025llw}, in which their mixing is a characteristic feature. This eLSM has also been applied to neutron-star physics~\cite{Ma:2025llw,Ma:2025tyt,Ma:2026rie}. Using this framework, we examine how meson mixing affects the nucleon D-form factor and compare its effect with that of the chiral invariant mass in the PDM. To obtain complementary information, we also study the pion D-form factor, which may help distinguish effects originating from the nucleon and meson sectors.




\section{Scalar meson dominance picture and the effect of chiral invariant mass}
In this section, we briefly review the scalar-meson-dominance picture of the nucleon $D$-form factor~\cite{Zwicky:2023fay,Kawaguchi:2025cuf}.
In this framework, the nucleon matrix element of the EMT is assumed to be saturated by scalar-meson exchanges. 
Retaining only the contribution from the lightest scalar meson, the nucleon $D$-form factor can be expressed in terms of its mass and the relevant couplings as
\begin{equation}
\begin{split}
D_N(t)= 
4m_N
\frac{
 g_{\sigma NN}\, g_{\sigma \Theta}
}{M^2_\sigma +t},
\label{general_DN}
\end{split}
\end{equation}
where $M_\sigma$ denotes the mass of the lightest sigma meson corresponding to $f_0(500)$, $g_{\sigma NN}$ denotes the Yukawa coupling between the sigma meson and the nucleon, and $g_{\sigma\Theta}$ denotes the coupling of the sigma meson to the EMT.

In the conventional two-flavor linear sigma model with the nucleon field, the meson fields are represented by the chiral field $\Phi = \sigma +i \sum_{a=1}^3\pi^a \tau^a$. In this model, the nucleon mass is generated entirely through spontaneous chiral symmetry breaking with the vacuum expectation value $\langle 0| \sigma | 0 \rangle =F_\pi$. Consequently, the Yukawa coupling is related to the nucleon mass and the pion decay constant as $g_{\sigma NN}^{\rm (LSM)} = m_N /F_\pi$. Meanwhile, the coupling $g_{\sigma\Theta}$ is determined by imposing the trace anomaly relation ${\Theta^\mu}_{\mu} = \partial_\mu j^\mu_D$,
where $j^\mu_D$ denotes the dilatation current. 
Evaluating the overlap amplitude between the improved EMT $\Theta^{\mu\nu}$ and the sigma meson state, one obtains $g_{\sigma \Theta}^{\rm (LSM)} = -F_\pi/3$.
Substituting these couplings into Eq.~\eqref{general_DN}, the nucleon $D$-form factor in the conventional LSM is obtained as~\cite{Zwicky:2023fay,Stegeman:2025sca,Kawaguchi:2025cuf}
\begin{equation}
D_N^{(\mathrm{LSM})}(t)
=
-\frac{4m_N^2}{3}
\frac{1}{t+M_\sigma^2}.
\label{ND_LSM}
\end{equation}
However, comparison with lattice QCD data~\cite{Hackett:2023rif} indicates that the LSM predicts a somewhat larger magnitude for the nucleon $D$-form factor.

To reduce the magnitude of the nucleon $D$-form factor, the previous study
considered the effect of a chiral invariant mass of the nucleon within the PDM~\cite{Kawaguchi:2025cuf}. In this model, the nucleon and its
negative-parity partner are introduced together with a
chiral-invariant mass $m_0$, which is independent of spontaneous chiral symmetry breaking~\cite{Detar:1988kn,Jido:2001nt}.
The Yukawa coupling is then modified as $g_{\sigma NN}^{\rm (PDM)} = m_N /F_\pi - 2m_0^2/\{F_\pi (m_N+m_{N^*})\}$ where $m_{N^*}$ denotes the mass of the negative-parity nucleon $N^*(1535)$.
Accordingly, the nucleon $D$-form factor in the PDM can be written as~\cite{Kawaguchi:2025cuf}
\begin{equation}
\begin{split}
D_N^{(\mathrm{PDM})}(t)
&=
C^{(m_0)}D_N^{(\mathrm{LSM})}(t),
\\
C^{(m_0)}
&=
1-
\frac{2m_0^2}{m_N(m_N + m_{N^*})}.
\label{ND_PDM}
\end{split}
\end{equation}
Thus, a nonzero value of $m_0$ suppresses the contribution from the lightest scalar meson exchange to the nucleon $D$-form factor, bringing it closer to the lattice-QCD result. 


\section{Extended linear sigma model}
We now examine whether a similar suppression can arise from the meson effect.
For this purpose, we employ the eLSM, which extends the conventional LSM by including both two-quark and four-quark states.
The inclusion of four-quark states is motivated by the long-standing possibility that light scalar mesons contain substantial four-quark components~\cite{Jaffe:1976ig,Pelaez:2015qba}.
A characteristic feature of the present model is the mixing between the two- and four-quark fields in both the scalar and pseudoscalar sectors. 
We therefore investigate how this meson mixing affects the
nucleon $D$-form factor.
In this section, we briefly review the eLSM and present the meson masses, the meson mixing structure, and the Yukawa couplings to the nucleon that
are relevant to the calculation of the nucleon $D$-form factor.

The eLSM with baryon fields is formulated for three quark flavors in Refs.~\cite{Ma:2023eoz,Ma:2025llw}. The model is constructed from two chiral nonet fields: one associated with a two-quark ($q\bar q$) state, denoted by $\Phi'$, and the other with a four-quark ($qq\bar q\bar q$) state, denoted by $\hat{\Phi}'$. In the linear representation, these fields are decomposed into scalar ($S$) and pseudoscalar ($P$) nonets as
\begin{equation}
\Phi'=S'+iP'
,\qquad
\hat\Phi'=\hat S'+i\hat P'
.
\end{equation}
Under $SU(3)_L\times SU(3)_R$ chiral transformations, $\Phi'$ and $\hat{\Phi}'$ transform in the same manner as the conventional chiral field $\Phi$, namely $\Phi \to g_L \Phi g_R^\dagger$ where $g_{L,R}\in SU(3)_{L,R}$.
However, under the $U(1)$ axial transformation, they carry different axial charges because of their different quark contents.

Using the building blocks $\Phi'$ and $\hat{\Phi}'$, the effective Lagrangian is expressed as
\begin{equation}
\mathcal{L}=\mathcal{L}_M+\mathcal{L}_B,
\end{equation}
where $\mathcal{L}_M$ and $\mathcal{L}_B$ denote the bosonic and baryonic sectors, respectively. The bosonic Lagrangian is given by
\begin{equation}
\begin{split}
\mathcal{L}_M
={}&
\frac{1}{2}\operatorname{Tr}
\left[
\partial_\mu\Phi'\partial^\mu\Phi'^\dagger
\right]
+
\frac{1}{2}\operatorname{Tr}
\left[
\partial_\mu\hat\Phi'\partial^\mu\hat\Phi'^\dagger
\right]
\\
&+
c_2\operatorname{Tr}\left[\Phi'\Phi'^\dagger\right]
-
d_2\operatorname{Tr}\left[\hat\Phi'\hat\Phi'^\dagger\right]
\\
&-
c_4\operatorname{Tr}
\left[
\Phi'\Phi'^\dagger\Phi'\Phi'^\dagger
\right]
\\
&-
e_3\left(
\epsilon_{abc}\epsilon_{def}
\Phi'_{ad}\Phi'_{be}\hat\Phi'_{cf}
+\mathrm{h.c.}
\right),
\label{eLSM_Lag}
\end{split}
\end{equation}
where $c_2$, $d_2$, $c_4$ and $e_3$ are coupling constants fixed in the previous study.

In the eLSM, spontaneous chiral symmetry breaking is realized through nonzero vacuum expectation values of the scalar states: $\langle0| S' |0\rangle=
\alpha\,{ 1}_{3\times3}$ and $\langle 0 | \hat S' |0\rangle
=
\beta\,{ 1}_{3\times3}$. These vacuum expectation values are related to the pion decay constant through 
$F_\pi^2=2(\alpha^2 + \beta^2)=(93\,{\rm MeV})^2$.
The resulting vacuum structure induces mixing between the isosinglet scalar components of the two-quark state, $\sigma'$, and the four-quark state, $\hat{\sigma}'$. This mixing is described by the following mass-squared matrix:
\begin{equation}
\mathcal{M}_S^2
=
\begin{pmatrix}
m_{\sigma'}^2 & m_{\sigma\mathrm{mix}}^2\\
m_{\sigma\mathrm{mix}}^2 & m_{\hat\sigma'}^2
\end{pmatrix},
\end{equation}
where  
$m_{\sigma'}^2 =8c_4 \alpha^2$,
$m_{\hat\sigma'}^2 = - 4 e_3 \alpha^2/\beta$, 
$m_{\sigma {\rm mix}}^2 = 8e_3 \alpha$.
In deriving these expressions, we have used the stationary conditions for $\alpha$ and $\beta$.
To diagonalize the mass matrix, we introduce the mass eigenstates by rotating $\sigma'$ and $\hat \sigma'$ with the mixing angle $\theta_S$:
\begin{equation}
\begin{pmatrix}
\hat \sigma\\
\sigma
\end{pmatrix}
=
\begin{pmatrix}
\cos\theta_S & \sin\theta_S\\
-\sin\theta_S & \cos\theta_S
\end{pmatrix}
\begin{pmatrix}
\sigma'\\
\hat\sigma'
\end{pmatrix},
\end{equation}
where $\sigma$ and $\hat \sigma$ denote the lightest and heavier scalar mass eigenstates, respectively. The corresponding mass eigenvalues are
\begin{equation}
\begin{split}
M_{\hat \sigma}^2
&= m_{\sigma'}^2 \cos^2\theta_S
+m_{\hat\sigma'}^2 \sin^2\theta_S
+2m_{\sigma {\rm mix}}^2\cos\theta_S \sin\theta_S
 ,\\
M_\sigma^2
&= m_{\sigma'}^2 \sin^2\theta_S
+m_{\hat\sigma'}^2 \cos^2\theta_S
-2m_{\sigma {\rm mix}}^2\cos\theta_S \sin\theta_S
,\\
\end{split}
\end{equation}
where $M_\sigma$ and $M_{\hat{\sigma}}$ correspond to the masses of $f_0(500)$ and $f_0(1500)$, respectively.

Mixing also occurs in the pseudoscalar sector. In particular, the physical isotriplet pseudoscalar mesons are the pion, $\pi$, and its excited state, $\hat{\pi}$. They are expressed in terms of the two-quark and four-quark basis states, $\pi'$ and $\hat{\pi}'$, through a mixing angle $\theta_P$:
\begin{equation}
\begin{split}
\begin{pmatrix}
\pi\\
\hat\pi
\end{pmatrix}
&=
\begin{pmatrix}
\cos \theta_P& \sin \theta_P\\
-\sin \theta_P& \cos \theta_P\\
\end{pmatrix}
\begin{pmatrix}
\pi'\\
\hat \pi'
\end{pmatrix}
\\
\end{split}
\end{equation}
In the present study, we work in the chiral limit. For the pion to be massless in the chiral limit, the mixing angle satisfies~\cite{Fariborz:2005gm,Fariborz:2007ai}
\begin{equation}
\begin{split}
\cos \theta_P&= \frac{\alpha}{\sqrt{\alpha^2 + \beta^2}},
\quad
\sin\theta_P
=
\frac{\beta}{\sqrt{\alpha^2+\beta^2}}.
\end{split}
\end{equation}

Turning to the baryonic sector, the interactions between the baryons and the scalar states are described by
\begin{equation}
\begin{split}
\mathcal{L}_B
={}&
\operatorname{Tr}
\left(\bar B i\gamma^\mu\partial_\mu B\right)
-
g\operatorname{Tr}\left(\bar B S'B\right)
\\
&+
e\epsilon_{abc}\epsilon_{def}
\bar B_{ad}B_{be}S'_{cf},
\end{split}
\end{equation}
where $B$ denotes the baryon octet field.
From this baryonic sector, the nucleon mass reads $m_N=(g+e)\alpha$, while the Yukawa couplings of the isosinglet scalar mesons to the nucleon are given by
\begin{equation}
\begin{split}
g_{\sigma NN}^{\rm (eLSM)} &
=
-\sqrt{\frac{2}{3}}\frac{m_N}{F_\pi} \frac{\sin\theta_S}{\cos \theta_P}
,
\quad
g_{\hat \sigma NN}^{\rm (eLSM)} 
=
\sqrt{\frac{2}{3}}\frac{m_N}{F_\pi} \frac{\cos\theta_S}{\cos \theta_P}
.
\end{split}
\end{equation}
Note that the factor $\sqrt{2/3}$ appearing here arises from the $\mathrm{U}(3)$-singlet normalization of $\sigma'$. A corresponding factor $\sqrt{3/2}$ will appear below in the coupling to the EMT. In a two-flavor model, such factors do not appear in the corresponding couplings. This difference originates from the fact that the $\sigma'$ field contains a strange component in the three-flavor model. However, this strange component does not directly contribute to the nucleon $D$-form factor in the scalar-meson-dominance picture. As a result, the normalization factors cancel in the product $g_{\sigma NN}g_{\sigma\Theta}$, which reproduces the corresponding two-flavor LSM result.

We next determine the couplings of the scalar mesons to the EMT. In the eLSM, the EMT is given by
\begin{equation}
\begin{split}
\Theta^{\mu\nu} &=
T^{\mu\nu} \\
&
-\frac{1}{6}(\partial^\mu \partial^\nu - g^{\mu\nu} \partial^2 ) 
\left(
{\rm tr}\left[
\Phi^{'} \Phi^{'\dagger} 
\right]
+
{\rm tr}\left[
\hat\Phi^{'} \hat\Phi^{'\dagger} 
\right]
\right),
\label{EMT_eLSM}
\end{split}
\end{equation}
where 
$T^{\mu\nu}$ denotes the Belinfante EMT derived from the eLSM Lagrangian. The second term in Eq.~\eqref{EMT_eLSM} is the improvement term~\cite{Callan:1970ze} introduced so that the trace of the EMT satisfies ${\Theta^\mu}_{\mu}=\partial_\mu j^\mu_D$.
After spontaneous chiral symmetry breaking, this term induces couplings of both the $\sigma$ and $\hat{\sigma}$ scalar fields to the EMT.
The corresponding couplings, $g_{\sigma\Theta}$ and
$g_{\hat{\sigma}\Theta}$, are obtained as
\begin{equation}
\begin{split}
g_{\sigma \Theta}^{\rm (eLSM)}
&=
\frac{1}{3}\sqrt{\frac{3}{2}} F_\pi \sin (\theta_S - \theta_P),
\\
g_{\hat \sigma \Theta}^{\rm (eLSM)}
&=
-\frac{1}{3}\sqrt{\frac{3}{2}} F_\pi \cos (\theta_S - \theta_P).
\end{split}
\end{equation}

\section{Nucleon D-form factor and four quark state/mixing structure}

Using the couplings derived above, we can now evaluate the nucleon $D$-form factor in the eLSM. Since both the lightest scalar meson, $\sigma$, and the
heavier scalar meson, $\hat{\sigma}$, couple to the nucleon and the EMT, their exchanges contribute to the nucleon $D$-form factor. The resulting form factor is expressed as
\begin{equation}
\begin{split}
&D_N^{\rm (eLSM)}(t)
=
4m_N
\frac{
 g_{\sigma NN}^{\rm (eLSM)}\, g_{\sigma \Theta}^{\rm (eLSM)}
}{M^2_\sigma +t}
+
4m_N
\frac{
 g_{\hat \sigma NN}^{\rm (eLSM)}\, g_{\hat \sigma \Theta}^{\rm (eLSM)}
}{M^2_{\hat \sigma} +t}
\\
&=
-\frac{4m_N^2}{3}
\Biggl[
\frac{\sin \theta_S}{\cos \theta_P} \sin(\theta_S -\theta_P)
\frac{1}{t+ M_{\sigma}^2}
\\
&
\qquad \quad
+
\frac{\cos \theta_S}{\cos \theta_P} \cos(\theta_S -\theta_P)
\frac{1}{t+M_{\hat \sigma}^2}
\Biggl].
\end{split}
\end{equation}
Since the contribution from the heavy scalar meson is suppressed by $1/M_{\hat{\sigma}}^2$, it can be safely neglected; its quantitative contribution is discussed in Appendix~\ref{app:sigmahat}.
The eLSM result
can then be expressed in terms of the LSM result multiplied by a factor:
\begin{equation}
D_N^{(\mathrm{eLSM})}(t)
=
C^{(\theta)}
D_N^{(\mathrm{LSM})}(t),
\end{equation}
where the factor $C^{(\theta)}$ depends on the mixing angles and is given by
\begin{equation}
C^{(\theta)}
=
\frac{\sin \theta_S}{\cos \theta_P} \sin(\theta_S -\theta_P)
.
\end{equation}
To determine the mixing angles numerically, we adopt the parameter set labeled el-g30g in the eLSM analysis of neutron stars in Ref.~\cite{Ma:2025llw}:
$\alpha=61.1\,{\rm MeV}$,
$\beta=24.7\,{\rm MeV}$,
$e_3=-2.06\,{\rm GeV}$,
$c_4 =44.0$,
$g = 15.4$,
and $e=-0.027$.
We have also checked the other parameter sets used in Ref.~\cite{Ma:2025llw}, and the corresponding results are listed in Appendix~\ref{app:sets}. The qualitative behavior is found to be insensitive to the choice of parameter set.
For the el-g30g parameter set, the resulting masses are
$m_N= 939.3\,{\rm MeV}$,
$M_\sigma = 521.8\,{\rm MeV}$,
$M_{\hat \sigma} = 1512.4\,{\rm MeV}$,
$M_\pi=0$,
and $M_{\hat\pi}=1204\,{\rm MeV}$.
Furthermore, the scalar and pseudoscalar mixing angles are obtained as $\theta_S = -44.0\tcdegree$ and
$\theta_P = 22.0\tcdegree$, which yields $C^{(\theta)} = 0.685$. 
This shows that meson mixing reduces the lightest-scalar meson contribution by about $30\%$.
Looking at this in more detail, $C^{(\theta)}$ can be rewritten as $C^{(\theta)}=\sin^2\theta_S-(\beta/\alpha)\sin\theta_S\cos\theta_S=0.483+0.202$.
The first term corresponds to the two-quark component of the physical $\sigma$, whereas the second term arises from the four-quark condensate through the factor $\beta/\alpha$.
Thus, the suppression is mainly associated with the two-quark component and is partially compensated by the four-quark condensate.

\begin{figure}
\includegraphics[scale=0.33]{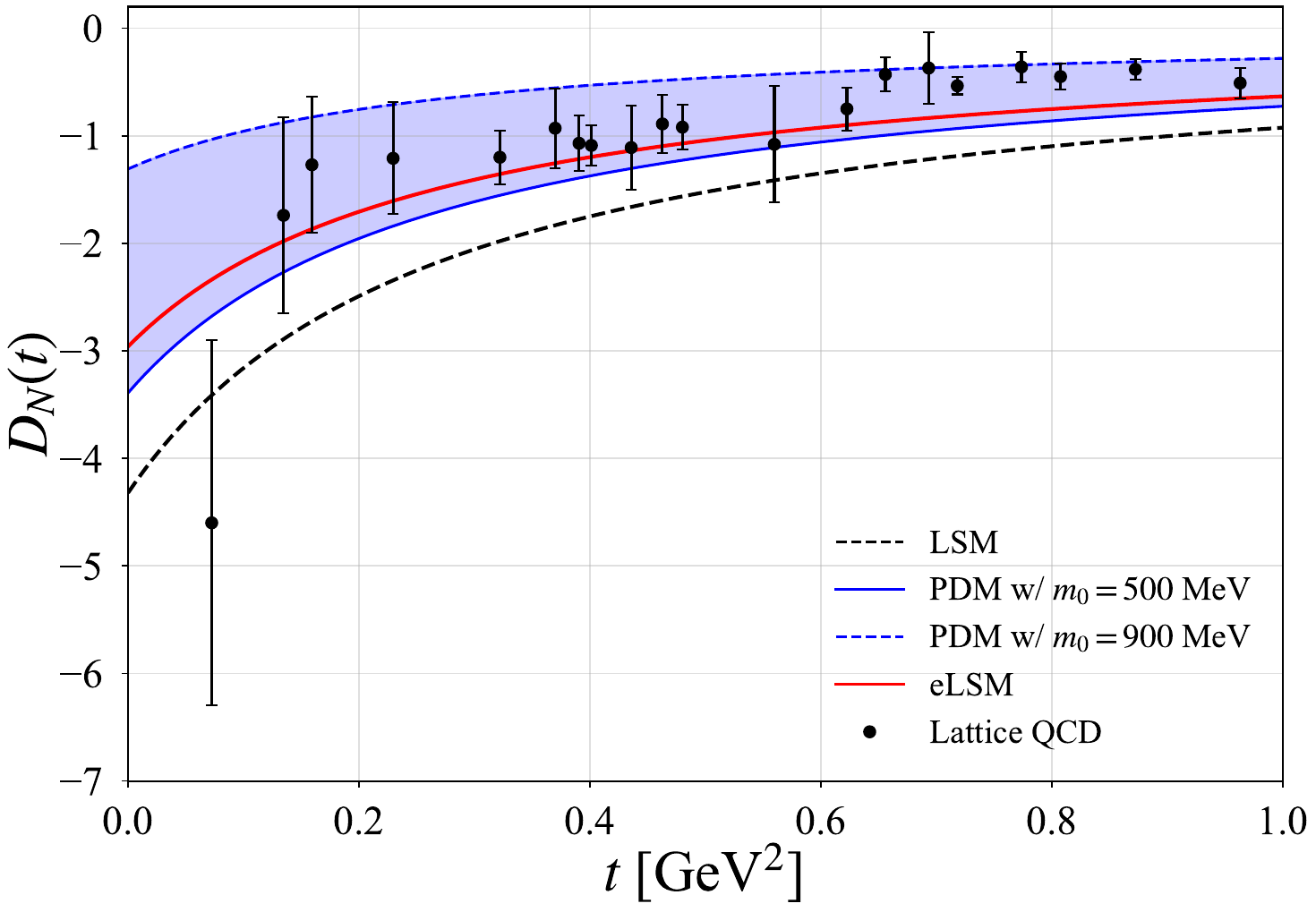}
\caption{Momentum-transfer dependence of the nucleon $D$-form factor in the LSM, PDM, and eLSM, compared with the lattice QCD data~\cite{Hackett:2023rif}. 
For the LSM and PDM calculations, the mass of the lightest scalar meson is fixed to the value obtained in the eLSM. Note that this choice should not be interpreted as the $\beta\to0$ limit of the eLSM.
 In the PDM, the chiral invariant mass is varied over the range $m_0=500\text{--}900\,{\rm MeV}$, consistent with constraints from neutron-star studies~\cite{Motohiro:2015taa,Minamikawa:2020jfj,Minamikawa:2023eky, Gao:2024lzu,Gao:2025nkg,Gao:2025vdc}, with $m_{N^*}=1535\,{\rm MeV}$.
}
  \label{NDform_eLSM}
\end{figure}

To clearly illustrate the momentum-transfer dependence of the nucleon $D$-form factor and compare the different approaches, we plot the eLSM result in Fig.~\ref{NDform_eLSM}, together with the LSM result in Eq.~\eqref{ND_LSM}, the PDM result in Eq.~\eqref{ND_PDM}, and the lattice QCD data~\cite{Hackett:2023rif}.
This figure shows that the LSM result deviates somewhat from the lattice data, whereas both the PDM results for $m_0=500\text{--}900\,{\rm MeV}$ and the eLSM result show better agreement with the lattice data. 
For $m_0=600\,{\rm MeV}$, the PDM suppression factor is $C^{(m_0)}=0.690$, which is nearly identical to the eLSM value $C^{(\theta)}=0.685$. This suggests that meson mixing and the chiral invariant mass are difficult to distinguish using the nucleon $D$-form factor alone.
Considering this situation, we next investigate the pion $D$-form factor, which is sensitive to meson mixing but not directly to the chiral-invariant mass, as a complementary observable that may help identify the origin of the suppression.

\section{Pion D-form factor in extended linear sigma model}

Analogously to the nucleon $D$-form factor, the pion $D$-form factor can also be expressed within the scalar meson dominance picture in terms of the coupling of the sigma meson to two pions, $g_{\sigma\pi\pi}$ with $\mathcal{L}_{\sigma\pi\pi}=-(1/2)g_{\sigma\pi\pi}\,\sigma\,\pi^a\pi^a$, and $g_{\sigma \Theta}$. 
A detailed derivation within the LSM can be found in Ref.~\cite{Kawaguchi:2026yae}.
In the LSM, it is given by
\begin{equation}
\begin{split}
D_\pi(t) 
= -\frac{1}{3} + 
\frac{2g_{\sigma\pi \pi}^{\rm (LSM)} g_{\sigma \Theta}^{\rm (LSM)}}{t+ M_{\sigma}^2}.
\end{split}
\end{equation}
In the chiral limit, the sigma-pion coupling is given by $g_{\sigma\pi \pi}^{\rm (LSM)} = M_\sigma^2/F_\pi$,
which can be readily obtained from the LSM sector employed in the PDM analysis~\cite{Kawaguchi:2025cuf}.
Using $g_{\sigma\Theta}^{\rm (LSM)}=-F_\pi/3$, the pion $D$-form factor becomes
\begin{equation}
\begin{split}
D_\pi^{(\text{LSM})}(t) 
=& -\frac{1}{3}
-
\frac{2}{3}
\frac{M_{\sigma}^2}{t+ M_{\sigma}^2}.
\label{piD_LSM}
\end{split}
\end{equation}
In the forward limit, this expression satisfies $D_\pi^{(\text{LSM})}(0) =-1$,
in agreement with the ChPT result~\cite{Voloshin:1982eb,Donoghue:1991qv}.
Since the mesonic sector of the PDM follows the LSM, the pion $D$-form factor in the PDM is identical to the LSM result in Eq.~\eqref{piD_LSM}.
This indicates that the chiral invariant mass does not directly affect the pion $D$-form factor.

In contrast, in the eLSM, the couplings of the isosinglet scalar mesons to pions are modified by the
mixing between mesons with the two-quark and four-quark states. From Lagrangian Eq.~\eqref{eLSM_Lag}, the corresponding couplings read 
\begin{equation}
\begin{split}
g_{\sigma \pi\pi}^{\rm (eLSM)}
&=-\sqrt{\frac{2}{3}}\frac{M_{\sigma}^2}{F_\pi}\sin(\theta_S-\theta_P),
\\
g_{\hat \sigma \pi\pi}^{\rm (eLSM)}
&=\sqrt{\frac{2}{3}}\frac{M_{\hat \sigma }^2}{F_\pi}
\cos(\theta_S-\theta_P).
\end{split}
\end{equation}
Using these couplings, the pion $D$-form factor in the eLSM is expressed as
\begin{equation}
\begin{split}
&D_\pi^{(\mathrm{eLSM})}(t)
=
-1 +\frac{2}{3}
+
2 \frac{
g_{\sigma \pi\pi}^{\rm (eLSM)}\, g_{\sigma \Theta}^{\rm (eLSM)}
}{M^2_\sigma +t}
\\
&\qquad\qquad
+
2 \frac{
g_{\hat \sigma \pi\pi}^{\rm (eLSM)}\, g_{\hat\sigma \Theta}^{\rm (eLSM)}
}{M^2_{\hat\sigma} +t}
\\
&=-\frac{1}{3}
-\frac{2}{3}\bigg[
\sin^2 (\theta_S -\theta_P) 
\frac{M_{\sigma}^2}{t+ M_{\sigma}^2}
\\
&\qquad\qquad
+
\cos^2(\theta_S-\theta_P) 
\frac{M_{\hat \sigma}^2}{t+ M_{\hat \sigma}^2}
\bigg]
\end{split}
\end{equation}
In contrast to the nucleon $D$-form factor, the contribution from the
heavier scalar meson cannot be neglected. This is because the $1/M_{\hat{\sigma}}^2$ suppression from its
propagator is canceled by the $M_{\hat{\sigma}}^2$ dependence of
$g_{\hat{\sigma}\pi\pi}^{\rm (eLSM)}$. 
The two scalar exchanges therefore preserve
\begin{equation}
\begin{split}
D_\pi^{(\mathrm{eLSM})}(0)
&=-1.
\end{split}
\end{equation}

\begin{figure}
\includegraphics[scale=0.33]{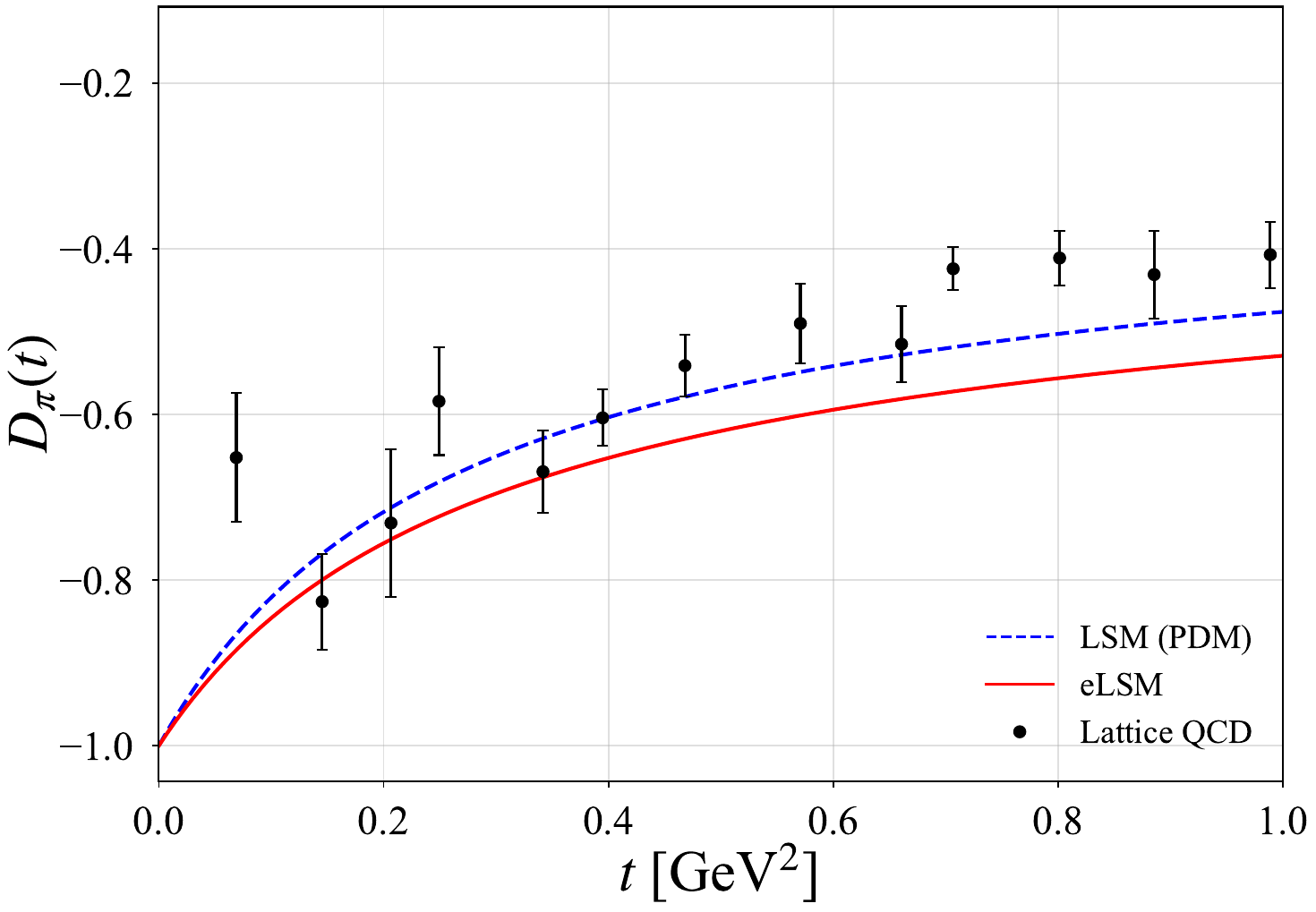}
\caption{Momentum-transfer dependence of the pion $D$-form factor in the LSM and eLSM, compared with the lattice-QCD results~\cite{Hackett:2023nkr}. The PDM result coincides with the LSM result due to their identical meson sectors. 
The effective model results are in the chiral limit, while the lattice-QCD results were obtained at $m_\pi \approx 170\,{\rm MeV}$.
}
  \label{piDform_eLSM}
\end{figure}

Nevertheless, meson mixing modifies the pion $D$-form factor at nonzero momentum transfer.
To illustrate this effect, Fig.~\ref{piDform_eLSM} shows the pion $D$-form factor in the eLSM using the parameter set adopted in the previous section, together with the LSM result and lattice-QCD results~\cite{Hackett:2023nkr} for comparison.
This figure shows that meson mixing enhances the magnitude of the pion $D$-form factor compared with the LSM result, and the deviation from the lattice-QCD result becomes more pronounced at larger momentum transfer. 
Note that in the eLSM result shown here, both the $\sigma$ and $\hat{\sigma}$ contributions are included.
However, a direct comparison should be made with care, since our effective-model result is obtained in the chiral limit, whereas the lattice-QCD result is evaluated at a finite pion mass.

To characterize this enhancement more quantitatively, we compare the slopes of the LSM and eLSM results at the forward limit, which are given by
\begin{equation}
\begin{split}
\frac{\partial D_\pi^{(\text{LSM})}(t)}{\partial t}\Biggl|_{t=0}&=
\frac{2}{3}
\frac{1}{ M_{\sigma}^2}
,\\
\frac{\partial D_\pi^{(\text{eLSM})}(t)}{\partial t}\Biggl|_{t=0}&=
\frac{2}{3}
\left[
\sin^2 (\theta_S -\theta_P) 
\frac{1}{M_{\sigma}^2}
\right.
\\
&\qquad\quad
\left.
+
\cos^2(\theta_S-\theta_P) 
\frac{1}{ M_{\hat \sigma}^2}
\right]
\\
&\simeq
\sin^2 (\theta_S -\theta_P) 
\frac{\partial D_\pi^{(\text{LSM})}(t)}{\partial t}\Biggl|_{t=0}
,
\end{split}
\end{equation}
In obtaining the last line, the heavier scalar-meson contribution to the slope has been neglected because of the $1/M_{\hat{\sigma}}^2$ suppression.
Since the mixing factor is $\sin^2(\theta_S-\theta_P)=0.835$, meson mixing reduces the slope of the pion $D$-form factor at the forward limit, thereby enhancing its magnitude at finite momentum transfer.

\section{Conclusion}

In this study, we investigated the gravitational $D$-form factors of the nucleon and pion in the sigma meson dominance picture using the eLSM that incorporates mixing between two-quark and four-quark states.
In particular, we examined whether meson mixing affects the nucleon $D$-form factor and how its effect differs from that of the chiral invariant mass found in our previous study.

We found that meson mixing suppresses the magnitude of the nucleon $D$-form factor. Quantitatively, for the parameter set adopted in this work, the suppression factor is $C^{(\theta)}=0.685$, leading to better agreement with the lattice-QCD result. 
Interestingly, this suppression is nearly identical to that obtained in the PDM with $m_0\simeq 600$ MeV.
Thus, although meson mixing and the chiral-invariant mass have physically different origins, their effects on the nucleon $D$-form factor are quantitatively very similar. This indicates that the nucleon $D$-form factor alone may not be sufficient to distinguish which mechanism underlies its behavior.

To obtain complementary information, we also studied the pion $D$-form factor. In the chiral limit, its forward value remains fixed at $D_\pi(0)=-1$ in both the LSM (PDM) and eLSM. However, meson mixing modifies its momentum-transfer dependence. In particular, the slope at the forward limit is reduced by the mixing effect, resulting in an enhancement of the magnitude of the pion $D$-form factor at finite momentum transfer compared with the LSM result. By contrast, the chiral invariant mass does not directly affect the pion $D$-form factor, as it does not enter the mesonic sector of the PDM.

We also compared the effective-model results for the pion $D$-form factor with the lattice-QCD data. However, the present analysis is performed in the chiral limit, whereas the lattice-QCD calculation is performed at a finite pion mass. A more direct comparison with the lattice-QCD results therefore requires the inclusion of explicit chiral-symmetry-breaking effects associated with the finite pion mass in the eLSM. Since such an extension makes the effective Lagrangian and the analysis more complex, we leave a detailed investigation at finite pion mass for future work.

These findings suggest that a combined analysis of the nucleon and pion $D$-form factors may provide a useful way to distinguish between effects originating from the baryon and meson sectors. In particular, while the nucleon $D$-form factor is sensitive to both the chiral structure of the nucleon mass and the meson mixing structure, the pion $D$-form factor provides more direct information on the mesonic sector. Such complementary information may therefore help clarify the origin of the nucleon mass as well as the quark composition and mixing structure of low-lying mesons.

Finally, we comment on a possible extension of the present analysis. In this study, we considered the scalar meson dominance picture, which is expected to provide a reasonable description in the low-energy region at small momentum transfer. At larger momentum transfer, contributions from heavier scalar mesons and tensor mesons may become important. The relevance of such contributions has recently been discussed in Refs.~\cite{Broniowski:2024oyk,Broniowski:2025ctl}. Including these additional meson contributions would therefore be a natural extension of the present analysis and will be considered in future work. 
Another possible extension is to include explicit chiral and flavor symmetry breaking, which may allow additional scalar-meson contributions to the nucleon $D$-form factor, motivated by the symmetry-breaking effects found in the LSM~\cite{Kawaguchi:2026yae}.

\section*{Acknowledgments}
The work of M. K. is supported by RFIS-NSFC under Grant No. W2433019.
The work of Y. M. is supported by Jiangsu Funding Program for Excellent Postdoctoral Talent under Grant No. 2025ZB516.
Y.~L. M. is supported in part by the National Science Foundation of China (NSFC) under Grant No. 12547104, the National Key R\&D Program of China under Grant No. 2021YFC2202900, and Gusu Talent Innovation Program under Grant No. ZXL2024363.

\appendix

\section{Contribution of the heavier scalar meson to the nucleon $D$-form factor}
\label{app:sigmahat}
Retaining both isosinglet scalar meson poles, the nucleon $D$-form factor is given by
\begin{equation}
D_N^{\rm (eLSM)}(t)=-\frac{4m_N^2}{3}
\left[\frac{C_\sigma^{(\theta)}}{t+M_\sigma^2}
+\frac{C_{\hat\sigma}^{(\theta)}}{t+M_{\hat\sigma}^2}\right],
\end{equation}
where
\begin{equation}
\begin{split}
C_\sigma^{(\theta)}&=\frac{\sin\theta_S\sin(\theta_S-\theta_P)}{\cos\theta_P},
\\
C_{\hat\sigma}^{(\theta)}&=\frac{\cos\theta_S\cos(\theta_S-\theta_P)}{\cos\theta_P},
\end{split}
\end{equation}
The coefficient $C^{(\theta)}$ introduced in the main text corresponds to $C_\sigma^{(\theta)}$. The two coefficients satisfy the exact sum rule
\begin{equation}
C_\sigma^{(\theta)}+C_{\hat\sigma}^{(\theta)}=1,
\label{eq:sumrule}
\end{equation}
which follows from 
\begin{equation}
\sum_{\phi=\sigma,\hat\sigma}g^{\rm (eLSM)}_{\phi NN}g^{\rm (eLSM)}_{\phi\Theta}=-\frac{m_N}{3}.
\end{equation}
Thus, the sum of the pole residues is fixed to the LSM normalization, and the meson mixing redistributes the strength between the two scalar poles. For the parameter set labeled el-g30g in Ref.~\cite{Ma:2025llw}, $C_\sigma^{(\theta)}=0.685$ and $C_{\hat\sigma}^{(\theta)}=0.315$. The relative size of the neglected $\hat{\sigma}$-pole contribution is defined as
\begin{equation}
R(t)=\frac{C_{\hat\sigma}^{(\theta)}\,(t+M_\sigma^2)}{C_\sigma^{(\theta)}\,(t+M_{\hat\sigma}^2)}.
\end{equation}
The resulting values are listed in Table~\ref{tab:Rt},
At the forward point, we obtain $D_N^{\rm (LSM)}(0)=-4.32$,
$D_N^{\rm (eLSM)}(0)=-2.96$ when only the $\sigma$ pole is retained, and $D_N^{\rm (eLSM)}(0)=-3.12$ when both scalar poles are included. Thus, including the $\hat\sigma$ pole shifts the eLSM result slightly toward the lattice-QCD data shown in Fig.~\ref{NDform_eLSM}.

Retaining the second pole does not spoil the suppression mechanism, but it changes its form: the constant factor $C^{(\theta)}$ is replaced by a momentum-dependent effective factor,
\begin{equation}
C^{(\theta)}_{\rm eff}(t)\equiv\frac{D_N^{\rm (eLSM)}(t)}{D_N^{\rm (LSM)}(t)}
=C_\sigma^{(\theta)}+C_{\hat\sigma}^{(\theta)}\,\frac{t+M_\sigma^2}{t+M_{\hat\sigma}^2}.
\label{eq:Ceff}
\end{equation}
At the forward point $C^{(\theta)}_{\rm eff}(0)=0.722$, corresponding to a suppression of about $28\%$. Thus, the suppression remains sizable after including the $\hat\sigma$ pole, and the agreement with the lattice-QCD data is maintained.
Formally, in the limit $t\to\infty$, $C_{\rm eff}^{(\theta)}(t)\to1$, using the sum rule in \eqref{eq:sumrule}.
Thus, within this expression, the eLSM result approaches the LSM result and the suppression due to meson mixing disappears at large \(t\). This suggests that the mixing effect is mainly relevant in the low-$t$ region.

The momentum dependence of $C^{(\theta)}_{\rm eff}(t)$
contrasts with the PDM, where the $t$-independent factor $C^{(m_0)}$ simply rescales the LSM form factor. To make a direct comparison, we choose $m_0=568\,{\rm MeV}$ such that the PDM reproduces $C_{\rm eff}^{(\theta)}(0)$. With this choice, the relative difference between the eLSM and PDM predictions increases with $t$, reaching $3.8\%$, $6.9\%$, and $11.7\%$ at $t=0.25$, $0.5$, and $1\,{\rm GeV}^2$, respectively.



\begin{table*}[htbp]
\caption{Relative size $R(t)$ of the $\hat\sigma$ contribution to $D_N^{\rm (eLSM)}(t)$, the resulting effective suppression factor $C^{(\theta)}_{\rm eff}(t)$ of Eq.~\eqref{eq:Ceff}, and the $t$-independent PDM factor $C^{(m_0)}$ with $m_0$ chosen to reproduce the eLSM result at $t=0$.  Results are shown for the parameter set labeled el-g30g.
}
\label{tab:Rt}
\centering
\begin{tabular}{lcccc}
\hline\hline
$t\,[{\rm GeV}^2]$ & 0 & 0.25 & 0.5 & 1.0 \\ 
\hline
$R(t)$ & $5.5\%$ & $9.5\%$ & $12.7\%$ & $17.8\%$ \\
$C^{(\theta)}_{\rm eff}(t)$ & 0.722 & 0.750 & 0.772 & 0.807 \\ 
$C^{(m_0)}$ w/ $m_0=568\,$MeV & 0.722 & 0.722 & 0.722 & 0.722 \\
\hline\hline
\end{tabular}
\end{table*}

\section{Dependence on the parameter set}
\label{app:sets}
Table~\ref{tab:sets} summarizes the meson masses, mixing angles and the mixing factors entering $D_N$ and $D_\pi$ for all six parameter sets considered in Ref.~\cite{Ma:2025llw}.
Across these sets, $C^{(\theta)}$ ranges from $0.672$ to $0.710$, while $\sin^2(\theta_S-\theta_P)$ ranges from  $0.832$ to $0.852$. These variations are only a few percent, indicating that the qualitative conclusions of the main text are insensitive to the choice of parameter set.
All parameter sets give scalar-meson masses consistent with $f_0(500)$ and $f_0(1500)$.

\begin{table*}[htbp]
\caption{
Model parameters, meson masses, and meson mixing angles for the six parameter sets considered in Ref.~\cite{Ma:2025llw}.
$\alpha$ and $\beta$ as well as meson masses are given in MeV, $e_3$ in GeV, and $c_4$ is dimensionless. 
$M_{\hat\pi}$, defined by $M_{\hat\pi}^2=-4e_3(\alpha^2+\beta^2)/\beta$, denots the mass of the excited pion in the chiral limit.
}
\label{tab:sets}
\centering
\small
\begin{tabular}{lcccccccccccc}
\hline\hline
set & $\alpha$ & $\beta$ & $e_3$ & $c_4$ & $m_N$ & $M_\sigma$ & $M_{\hat\sigma}$ & $M_{\hat\pi}$ & $\theta_S$ & $\theta_P$ & $C^{(\theta)}$ & $\sin^2(\theta_S-\theta_P)$ \\
\hline
el-g30g    & 61.1 & 24.7 & $-2.06$ & 44.0 & 939.3 & 521.8 & 1512.4 & 1203.7 & $-44.02^\circ$ & $22.01^\circ$ & 0.685 & 0.835 \\
el-g30e    & 61.1 & 24.4 & $-2.05$ & 43.6 & 936.9 & 525.4 & 1510.2 & 1206.1 & $-44.32^\circ$ & $21.77^\circ$ & 0.688 & 0.836 \\
el-g30eg   & 61.4 & 26.4 & $-2.10$ & 45.6 & 938.8 & 502.1 & 1524.0 & 1192.2 & $-42.57^\circ$ & $23.27^\circ$ & 0.672 & 0.832 \\
el-g350eg  & 61.2 & 25.6 & $-2.10$ & 44.4 & 939.4 & 500.3 & 1519.6 & 1201.7 & $-43.59^\circ$ & $22.70^\circ$ & 0.684 & 0.838 \\
el-g3100eg & 60.8 & 24.0 & $-2.09$ & 42.4 & 938.8 & 504.0 & 1512.4 & 1220.0 & $-45.48^\circ$ & $21.54^\circ$ & 0.706 & 0.848 \\
el-g3150eg & 60.7 & 24.3 & $-2.11$ & 42.0 & 937.8 & 483.8 & 1511.2 & 1218.5 & $-45.58^\circ$ & $21.82^\circ$ & 0.710 & 0.852 \\
\hline\hline
\end{tabular}
\end{table*}

\bibliographystyle{apsrev4-1}
\bibliography{reference}

@article{Gao:2025vdc,
    author = "Gao, Bikai and Kong, Yuk-Kei and Ma, Yong-Liang",
    title = "{Origin of nucleon mass in the light of PSR J0614-3329 with quark-hadron crossover}",
    eprint = "2509.03008",
    archivePrefix = "arXiv",
    primaryClass = "nucl-th",
    doi = "10.1103/xwkv-s8lg",
    journal = "Phys. Rev. D",
    volume = "112",
    number = "8",
    pages = "083041",
    year = "2025"
}

@article{Gao:2025nkg,
    author = "Gao, Bikai and Liu, Xiang and Harada, Masayasu and Ma, Yong-Liang",
    title = "{Implication of neutron star observations to the origin of nucleon mass}",
    eprint = "2508.00243",
    archivePrefix = "arXiv",
    primaryClass = "nucl-th",
    doi = "10.1007/s11433-025-2839-7",
    journal = "Sci. China Phys. Mech. Astron.",
    volume = "69",
    number = "3",
    pages = "232011",
    year = "2026"
}

@article{Hackett:2023rif,
    author = "Hackett, Daniel C. and Pefkou, Dimitra A. and Shanahan, Phiala E.",
    title = "{Gravitational Form Factors of the Proton from Lattice QCD}",
    eprint = "2310.08484",
    archivePrefix = "arXiv",
    primaryClass = "hep-lat",
    reportNumber = "MIT-CTP/5630, FERMILAB-PUB-23-592-T",
    doi = "10.1103/PhysRevLett.132.251904",
    journal = "Phys. Rev. Lett.",
    volume = "132",
    number = "25",
    pages = "251904",
    year = "2024"
}

@article{Stegeman:2025tdl,
    author = "Stegeman, Roy and Zwicky, Roman",
    title = "{Gluon gravitational D-form factor: the {\ensuremath{\sigma}}-meson as a dilaton confronted with lattice data II}",
    eprint = "2512.12315",
    archivePrefix = "arXiv",
    primaryClass = "hep-ph",
    doi = "10.1007/JHEP05(2026)159",
    journal = "JHEP",
    volume = "05",
    pages = "159",
    year = "2026"
}

@article{Stegeman:2025sca,
    author = "Stegeman, Roy and Zwicky, Roman",
    title = "{Gravitational D-form factor: the {\ensuremath{\sigma}}-meson as a dilaton confronted with lattice QCD data I}",
    eprint = "2508.18537",
    archivePrefix = "arXiv",
    primaryClass = "hep-ph",
    doi = "10.1007/JHEP03(2026)184",
    journal = "JHEP",
    volume = "03",
    pages = "184",
    year = "2026"
}

@article{Detar:1988kn,
    author = "Detar, Carleton E. and Kunihiro, Teiji",
    title = "{Linear $\sigma$ Model With Parity Doubling}",
    reportNumber = "RIFP-748",
    doi = "10.1103/PhysRevD.39.2805",
    journal = "Phys. Rev. D",
    volume = "39",
    pages = "2805",
    year = "1989"
}

@article{Jido:2001nt,
    author = "Jido, Daisuke and Oka, Makoto and Hosaka, Atsushi",
    title = "{Chiral symmetry of baryons}",
    eprint = "hep-ph/0110005",
    archivePrefix = "arXiv",
    doi = "10.1143/PTP.106.873",
    journal = "Prog. Theor. Phys.",
    volume = "106",
    pages = "873--908",
    year = "2001"
}

@misc{Burkert:2021ith,
    author = "Burkert, V. D. and Elouadrhiri, L. and Girod, F. X.",
    title = "{Determination of shear forces inside the proton}",
    eprint = "2104.02031",
    archivePrefix = "arXiv",
    primaryClass = "nucl-ex",
    month = "4",
    year = "2021"
}

@article{Broniowski:2024oyk,
    author = "Broniowski, Wojciech and Ruiz Arriola, Enrique",
    title = "{Gravitational form factors of the pion and meson dominance}",
    eprint = "2405.07815",
    archivePrefix = "arXiv",
    primaryClass = "hep-ph",
    doi = "10.1016/j.physletb.2024.139138",
    journal = "Phys. Lett. B",
    volume = "859",
    pages = "139138",
    year = "2024"
}

@article{Motohiro:2015taa,
    author = "Motohiro, Yuichi and Kim, Youngman and Harada, Masayasu",
    title = "{Asymmetric nuclear matter in a parity doublet model with hidden local symmetry}",
    eprint = "1505.00988",
    archivePrefix = "arXiv",
    primaryClass = "nucl-th",
    doi = "10.1103/PhysRevC.92.025201",
    journal = "Phys. Rev. C",
    volume = "92",
    number = "2",
    pages = "025201",
    year = "2015",
    note = "[Erratum: Phys.Rev.C 95, 059903 (2017)]"
}

@article{Burkert:2018bqq,
    author = "Burkert, V. D. and Elouadrhiri, L. and Girod, F. X.",
    title = "{The pressure distribution inside the proton}",
    doi = "10.1038/s41586-018-0060-z",
    journal = "Nature",
    volume = "557",
    number = "7705",
    pages = "396--399",
    year = "2018"
}

@article{Duran:2022xag,
    author = "Duran, B. and others",
    title = "{Determining the gluonic gravitational form factors of the proton}",
    eprint = "2207.05212",
    archivePrefix = "arXiv",
    primaryClass = "nucl-ex",
    doi = "10.1038/s41586-023-05730-4",
    journal = "Nature",
    volume = "615",
    number = "7954",
    pages = "813--816",
    year = "2023"
}

@article{Polyakov:2018exb,
    author = "Polyakov, Maxim V. and Son, Hyeon-Dong",
    title = "{Nucleon gravitational form factors from instantons: forces between quark and gluon subsystems}",
    eprint = "1808.00155",
    archivePrefix = "arXiv",
    primaryClass = "hep-ph",
    doi = "10.1007/JHEP09(2018)156",
    journal = "JHEP",
    volume = "09",
    pages = "156",
    year = "2018"
}

@article{Shanahan:2018nnv,
    author = "Shanahan, P. E. and Detmold, W.",
    title = "{Pressure Distribution and Shear Forces inside the Proton}",
    eprint = "1810.07589",
    archivePrefix = "arXiv",
    primaryClass = "nucl-th",
    reportNumber = "MIT-CTP/5071",
    doi = "10.1103/PhysRevLett.122.072003",
    journal = "Phys. Rev. Lett.",
    volume = "122",
    number = "7",
    pages = "072003",
    year = "2019"
}

@article{Lorce:2018egm,
    author = "Lorc\'e, C\'edric and Moutarde, Herv\'e and Trawi\'nski, Arkadiusz P.",
    title = "{Revisiting the mechanical properties of the nucleon}",
    eprint = "1810.09837",
    archivePrefix = "arXiv",
    primaryClass = "hep-ph",
    doi = "10.1140/epjc/s10052-019-6572-3",
    journal = "Eur. Phys. J. C",
    volume = "79",
    number = "1",
    pages = "89",
    year = "2019"
}

@article{Anikin:2019kwi,
    author = "Anikin, I. V.",
    title = "{Gravitational form factors within light-cone sum rules at leading order}",
    eprint = "1902.00094",
    archivePrefix = "arXiv",
    primaryClass = "hep-ph",
    doi = "10.1103/PhysRevD.99.094026",
    journal = "Phys. Rev. D",
    volume = "99",
    number = "9",
    pages = "094026",
    year = "2019"
}

@article{Avelino:2019esh,
    author = "Avelino, P. P.",
    title = "{Probing gravity at sub-femtometer scales through the pressure distribution inside the proton}",
    eprint = "1902.01318",
    archivePrefix = "arXiv",
    primaryClass = "gr-qc",
    doi = "10.1016/j.physletb.2019.05.056",
    journal = "Phys. Lett. B",
    volume = "795",
    pages = "627--631",
    year = "2019"
}

@article{Yanagihara:2019foh,
    author = "Yanagihara, Ryosuke and Kitazawa, Masakiyo",
    title = "{A study of stress-tensor distribution around the flux tube in the Abelian\textendash{}Higgs model}",
    eprint = "1905.10056",
    archivePrefix = "arXiv",
    primaryClass = "hep-ph",
    reportNumber = "J-PARC-TH-0165",
    doi = "10.1093/ptep/ptz093",
    journal = "PTEP",
    volume = "2019",
    number = "9",
    pages = "093B02",
    year = "2019",
    note = "[Erratum: PTEP 2020, 079201 (2020)]"
}

@article{Hatta:2019lxo,
    author = "Hatta, Yoshitaka and Rajan, Abha and Yang, Di-Lun",
    title = "{Near threshold J/\ensuremath{\psi} and \ensuremath{\Upsilon} photoproduction at JLab and RHIC}",
    eprint = "1906.00894",
    archivePrefix = "arXiv",
    primaryClass = "hep-ph",
    reportNumber = "YITP-19-46",
    doi = "10.1103/PhysRevD.100.014032",
    journal = "Phys. Rev. D",
    volume = "100",
    number = "1",
    pages = "014032",
    year = "2019"
}

@article{Freese:2019eww,
    author = {Freese, Adam and Clo\"et, Ian C.},
    title = "{Impact of dynamical chiral symmetry breaking and dynamical diquark correlations on proton generalized parton distributions}",
    eprint = "1907.08256",
    archivePrefix = "arXiv",
    primaryClass = "nucl-th",
    doi = "10.1103/PhysRevC.101.035203",
    journal = "Phys. Rev. C",
    volume = "101",
    number = "3",
    pages = "035203",
    year = "2020"
}

@article{Azizi:2019ytx,
    author = {Azizi, K. and \"Ozdem, U.},
    title = "{Nucleon\textquoteright{}s energy\textendash{}momentum tensor form factors in light-cone QCD}",
    eprint = "1908.06143",
    archivePrefix = "arXiv",
    primaryClass = "hep-ph",
    doi = "10.1140/epjc/s10052-020-7676-5",
    journal = "Eur. Phys. J. C",
    volume = "80",
    number = "2",
    pages = "104",
    year = "2020"
}

@article{Mamo:2019mka,
    author = "Mamo, Kiminad A. and Zahed, Ismail",
    title = "{Diffractive photoproduction of $J/\psi$ and $\Upsilon$ using holographic QCD: gravitational form factors and GPD of gluons in the proton}",
    eprint = "1910.04707",
    archivePrefix = "arXiv",
    primaryClass = "hep-ph",
    doi = "10.1103/PhysRevD.101.086003",
    journal = "Phys. Rev. D",
    volume = "101",
    number = "8",
    pages = "086003",
    year = "2020"
}

@article{Neubelt:2019sou,
    author = "Neubelt, Matt J. and Sampino, Andrew and Hudson, Jonathan and Tezgin, Kemal and Schweitzer, Peter",
    title = "{Energy momentum tensor and the D-term in the bag model}",
    eprint = "1911.08906",
    archivePrefix = "arXiv",
    primaryClass = "hep-ph",
    doi = "10.1103/PhysRevD.101.034013",
    journal = "Phys. Rev. D",
    volume = "101",
    number = "3",
    pages = "034013",
    year = "2020"
}

@article{Alharazin:2020yjv,
    author = "Alharazin, H. and Djukanovic, D. and Gegelia, J. and Polyakov, M. V.",
    title = "{Chiral theory of nucleons and pions in the presence of an external gravitational field}",
    eprint = "2006.05890",
    archivePrefix = "arXiv",
    primaryClass = "hep-ph",
    doi = "10.1103/PhysRevD.102.076023",
    journal = "Phys. Rev. D",
    volume = "102",
    number = "7",
    pages = "076023",
    year = "2020"
}

@article{Varma:2020crx,
    author = "Varma, Mira and Schweitzer, Peter",
    title = "{Effects of long-range forces on the D-term and the energy-momentum structure}",
    eprint = "2006.06602",
    archivePrefix = "arXiv",
    primaryClass = "hep-ph",
    doi = "10.1103/PhysRevD.102.014047",
    journal = "Phys. Rev. D",
    volume = "102",
    number = "1",
    pages = "014047",
    year = "2020"
}

@article{Kim:2020nug,
    author = "Kim, June-Young and Kim, Hyun-Chul and Polyakov, Maxim V. and Son, Hyeon-Dong",
    title = "{Strong force fields and stabilities of the nucleon and singly heavy baryon $\Sigma_c$}",
    eprint = "2008.06652",
    archivePrefix = "arXiv",
    primaryClass = "hep-ph",
    reportNumber = "INHA-NTG-04/2020",
    doi = "10.1103/PhysRevD.103.014015",
    journal = "Phys. Rev. D",
    volume = "103",
    number = "1",
    pages = "014015",
    year = "2021"
}

@article{Chakrabarti:2020kdc,
    author = "Chakrabarti, Dipankar and Mondal, Chandan and Mukherjee, Asmita and Nair, Sreeraj and Zhao, Xingbo",
    title = "{Gravitational form factors and mechanical properties of proton in a light-front quark-diquark model}",
    eprint = "2010.04215",
    archivePrefix = "arXiv",
    primaryClass = "hep-ph",
    doi = "10.1103/PhysRevD.102.113011",
    journal = "Phys. Rev. D",
    volume = "102",
    pages = "113011",
    year = "2020"
}

@article{Yanagihara:2020tvs,
    author = "Yanagihara, Ryosuke and Kitazawa, Masakiyo and Asakawa, Masayuki and Hatsuda, Tetsuo",
    title = "{Distribution of Energy-Momentum Tensor around a Static Quark in the Deconfined Phase of SU(3) Yang-Mills Theory}",
    eprint = "2010.13465",
    archivePrefix = "arXiv",
    primaryClass = "hep-lat",
    reportNumber = "J-PARC-TH-0229, RIKEN-iTHEMS-Report-20",
    doi = "10.1103/PhysRevD.102.114522",
    journal = "Phys. Rev. D",
    volume = "102",
    number = "11",
    pages = "114522",
    year = "2020"
}

@article{Kim:2020lrs,
    author = "Kim, June-Young and Sun, Bao-Dong",
    title = "{Gravitational form factors of a baryon with spin-3/2}",
    eprint = "2011.00292",
    archivePrefix = "arXiv",
    primaryClass = "hep-ph",
    doi = "10.1140/epjc/s10052-021-08852-z",
    journal = "Eur. Phys. J. C",
    volume = "81",
    number = "1",
    pages = "85",
    year = "2021"
}

@article{Tong:2021ctu,
    author = "Tong, Xuan-Bo and Ma, Jian-Ping and Yuan, Feng",
    title = "{Gluon gravitational form factors at large momentum transfer}",
    eprint = "2101.02395",
    archivePrefix = "arXiv",
    primaryClass = "hep-ph",
    doi = "10.1016/j.physletb.2021.136751",
    journal = "Phys. Lett. B",
    volume = "823",
    pages = "136751",
    year = "2021"
}

@article{Freese:2021czn,
    author = "Freese, Adam and Miller, Gerald A.",
    title = "{Forces within hadrons on the light front}",
    eprint = "2102.01683",
    archivePrefix = "arXiv",
    primaryClass = "hep-ph",
    reportNumber = "NT@UW-21-02",
    doi = "10.1103/PhysRevD.103.094023",
    journal = "Phys. Rev. D",
    volume = "103",
    pages = "094023",
    year = "2021"
}

@article{Panteleeva:2021iip,
    author = "Panteleeva, Julia Yu. and Polyakov, Maxim V.",
    title = "{Forces inside the nucleon on the light front from 3D Breit frame force distributions: Abel tomography case}",
    eprint = "2102.10902",
    archivePrefix = "arXiv",
    primaryClass = "hep-ph",
    doi = "10.1103/PhysRevD.104.014008",
    journal = "Phys. Rev. D",
    volume = "104",
    number = "1",
    pages = "014008",
    year = "2021"
}

@article{Hatta:2021can,
    author = "Hatta, Yoshitaka and Strikman, Mark",
    title = "{$\phi$-meson lepto-production near threshold and the strangeness $D$-term}",
    eprint = "2102.12631",
    archivePrefix = "arXiv",
    primaryClass = "hep-ph",
    doi = "10.1016/j.physletb.2021.136295",
    journal = "Phys. Lett. B",
    volume = "817",
    pages = "136295",
    year = "2021"
}

@article{Mamo:2021krl,
    author = "Mamo, Kiminad A. and Zahed, Ismail",
    title = "{Nucleon mass radii and distribution: Holographic QCD, Lattice QCD and GlueX data}",
    eprint = "2103.03186",
    archivePrefix = "arXiv",
    primaryClass = "hep-ph",
    doi = "10.1103/PhysRevD.103.094010",
    journal = "Phys. Rev. D",
    volume = "103",
    number = "9",
    pages = "094010",
    year = "2021"
}

@article{Freese:2021qtb,
    author = "Freese, Adam and Miller, Gerald A.",
    title = "{Genuine empirical pressure within the proton}",
    eprint = "2104.03213",
    archivePrefix = "arXiv",
    primaryClass = "hep-ph",
    reportNumber = "NT@UW-21-03",
    doi = "10.1103/PhysRevD.104.014024",
    journal = "Phys. Rev. D",
    volume = "104",
    number = "1",
    pages = "014024",
    year = "2021"
}

@article{Gegelia:2021wnj,
    author = "Gegelia, Jambul and Polyakov, Maxim V.",
    title = "{A bound on the nucleon Druck-term from chiral EFT in curved space-time and mechanical stability conditions}",
    eprint = "2104.13954",
    archivePrefix = "arXiv",
    primaryClass = "hep-ph",
    doi = "10.1016/j.physletb.2021.136572",
    journal = "Phys. Lett. B",
    volume = "820",
    pages = "136572",
    year = "2021"
}

@article{Kim:2021jjf,
    author = "Kim, June-Young and Kim, Hyun-Chul",
    title = "{Energy-momentum tensor of the nucleon on the light front: Abel tomography case}",
    eprint = "2105.10279",
    archivePrefix = "arXiv",
    primaryClass = "hep-ph",
    reportNumber = "INHA-NTG-04/2021",
    doi = "10.1103/PhysRevD.104.074019",
    journal = "Phys. Rev. D",
    volume = "104",
    number = "7",
    pages = "074019",
    year = "2021"
}

@article{Owa:2021hnj,
    author = "Owa, Shiryo and Thomas, A. W. and Wang, X. G.",
    title = "{Effect of the pion field on the distributions of pressure and shear in the proton}",
    eprint = "2106.00929",
    archivePrefix = "arXiv",
    primaryClass = "hep-ph",
    reportNumber = "ADP-21-8/T1156",
    doi = "10.1016/j.physletb.2022.137136",
    journal = "Phys. Lett. B",
    volume = "829",
    pages = "137136",
    year = "2022"
}

@article{Pefkou:2021fni,
    author = "Pefkou, Dimitra A. and Hackett, Daniel C. and Shanahan, Phiala E.",
    title = "{Gluon gravitational structure of hadrons of different spin}",
    eprint = "2107.10368",
    archivePrefix = "arXiv",
    primaryClass = "hep-lat",
    reportNumber = "MIT-CTP/5318",
    doi = "10.1103/PhysRevD.105.054509",
    journal = "Phys. Rev. D",
    volume = "105",
    number = "5",
    pages = "054509",
    year = "2022"
}

@article{Lorce:2021xku,
    author = "Lorc\'e, C\'edric and Metz, Andreas and Pasquini, Barbara and Rodini, Simone",
    title = "{Energy-momentum tensor in QCD: nucleon mass decomposition and mechanical equilibrium}",
    eprint = "2109.11785",
    archivePrefix = "arXiv",
    primaryClass = "hep-ph",
    doi = "10.1007/JHEP11(2021)121",
    journal = "JHEP",
    volume = "11",
    pages = "121",
    year = "2021"
}

@article{Ji:2021mfb,
    author = "Ji, Xiangdong and Liu, Yizhuang",
    title = "{Momentum-Current Gravitational Multipoles of Hadrons}",
    eprint = "2110.14781",
    archivePrefix = "arXiv",
    primaryClass = "hep-ph",
    doi = "10.1103/PhysRevD.106.034028",
    journal = "Phys. Rev. D",
    volume = "106",
    number = "3",
    pages = "034028",
    year = "2022"
}

@article{More:2021stk,
    author = "More, Jai and Mukherjee, Asmita and Nair, Sreeraj and Saha, Sudeep",
    title = "{Gravitational form factors and mechanical properties of a quark at one loop in light-front Hamiltonian QCD}",
    eprint = "2112.06550",
    archivePrefix = "arXiv",
    primaryClass = "hep-ph",
    doi = "10.1103/PhysRevD.105.056017",
    journal = "Phys. Rev. D",
    volume = "105",
    number = "5",
    pages = "056017",
    year = "2022"
}

@article{Mamo:2022eui,
    author = "Mamo, Kiminad A. and Zahed, Ismail",
    title = "{J/\ensuremath{\psi} near threshold in holographic QCD: A and D gravitational form factors}",
    eprint = "2204.08857",
    archivePrefix = "arXiv",
    primaryClass = "hep-ph",
    doi = "10.1103/PhysRevD.106.086004",
    journal = "Phys. Rev. D",
    volume = "106",
    number = "8",
    pages = "086004",
    year = "2022"
}

@article{Lorce:2022cle,
    author = "Lorc\'e, C\'edric and Schweitzer, Peter and Tezgin, Kemal",
    title = "{2D energy-momentum tensor distributions of nucleon in a large-Nc quark model from ultrarelativistic to nonrelativistic limit}",
    eprint = "2202.01192",
    archivePrefix = "arXiv",
    primaryClass = "hep-ph",
    doi = "10.1103/PhysRevD.106.014012",
    journal = "Phys. Rev. D",
    volume = "106",
    number = "1",
    pages = "014012",
    year = "2022"
}

@article{Fujita:2022jus,
    author = "Fujita, Mitsutoshi and Hatta, Yoshitaka and Sugimoto, Shigeki and Ueda, Takahiro",
    title = "{Nucleon D-term in holographic quantum chromodynamics}",
    eprint = "2206.06578",
    archivePrefix = "arXiv",
    primaryClass = "hep-th",
    reportNumber = "YITP-22-08, KUNS-2931",
    doi = "10.1093/ptep/ptac110",
    journal = "PTEP",
    volume = "2022",
    number = "9",
    pages = "093B06",
    year = "2022"
}

@article{Choudhary:2022den,
    author = "Choudhary, Poonam and Gurjar, Bheemsehan and Chakrabarti, Dipankar and Mukherjee, Asmita",
    title = "{Gravitational form factors and mechanical properties of the proton: Connections between distributions in 2D and 3D}",
    eprint = "2206.12206",
    archivePrefix = "arXiv",
    primaryClass = "hep-ph",
    doi = "10.1103/PhysRevD.106.076004",
    journal = "Phys. Rev. D",
    volume = "106",
    number = "7",
    pages = "076004",
    year = "2022"
}

@article{Kim:2022wkc,
    author = "Kim, June-Young and Sun, Bao-Dong and Fu, Dongyan and Kim, Hyun-Chul",
    title = "{Mechanical structure of a spin-1 particle}",
    eprint = "2208.01240",
    archivePrefix = "arXiv",
    primaryClass = "hep-ph",
    reportNumber = "INHA-NTG-06/2022",
    doi = "10.1103/PhysRevD.107.054007",
    journal = "Phys. Rev. D",
    volume = "107",
    number = "5",
    pages = "054007",
    year = "2023"
}

@article{Alharazin:2022wjj,
    author = "Alharazin, H. and Epelbaum, E. and Gegelia, J. and Mei\ss{}ner, U. -G. and Sun, B. -D.",
    title = "{Gravitational form factors of the delta resonance in chiral EFT}",
    eprint = "2209.01233",
    archivePrefix = "arXiv",
    primaryClass = "hep-ph",
    doi = "10.1140/epjc/s10052-022-10882-0",
    journal = "Eur. Phys. J. C",
    volume = "82",
    number = "10",
    pages = "907",
    year = "2022"
}

@article{Won:2022cyy,
    author = "Won, Ho-Yeon and Kim, June-Young and Kim, Hyun-Chul",
    title = "{Gravitational form factors of the baryon octet with flavor SU(3) symmetry breaking}",
    eprint = "2210.03320",
    archivePrefix = "arXiv",
    primaryClass = "hep-ph",
    reportNumber = "INHA-NTG-11/2022",
    doi = "10.1103/PhysRevD.106.114009",
    journal = "Phys. Rev. D",
    volume = "106",
    number = "11",
    pages = "114009",
    year = "2022"
}

@article{Tanaka:2022wzy,
    author = "Tanaka, Kazuhiro",
    title = "{Twist-four gravitational form factor at NNLO QCD from trace anomaly constraints}",
    eprint = "2212.09417",
    archivePrefix = "arXiv",
    primaryClass = "hep-ph",
    doi = "10.1007/JHEP03(2023)013",
    journal = "JHEP",
    volume = "03",
    pages = "013",
    year = "2023"
}

@article{Ito:2023oby,
    author = "Ito, Hiroaki and Kitazawa, Masakiyo",
    title = "{Gravitational form factors of a kink in 1 + 1 dimensional \ensuremath{\phi}$^{4}$ model}",
    eprint = "2302.08762",
    archivePrefix = "arXiv",
    primaryClass = "hep-th",
    reportNumber = "YITP-23-18, J-PARC-TH-0284",
    doi = "10.1007/JHEP08(2023)033",
    journal = "JHEP",
    volume = "08",
    pages = "033",
    year = "2023"
}

@article{Lorce:2023zzg,
    author = "Lorc\'e, C\'edric and Song, Qin-Tao",
    title = "{Gravitational transverse-momentum distributions}",
    eprint = "2303.11538",
    archivePrefix = "arXiv",
    primaryClass = "hep-ph",
    doi = "10.1016/j.physletb.2023.138016",
    journal = "Phys. Lett. B",
    volume = "843",
    pages = "138016",
    year = "2023"
}

@article{Amor-Quiroz:2023rke,
    author = "Amor-Quiroz, Arturo and Focillon, William and Lorc\'e, C\'edric and Rodini, Simone",
    title = "{Energy\textendash{}momentum tensor in the scalar diquark model}",
    eprint = "2304.10339",
    archivePrefix = "arXiv",
    primaryClass = "hep-ph",
    doi = "10.1140/epjc/s10052-023-12190-7",
    journal = "Eur. Phys. J. C",
    volume = "83",
    number = "11",
    pages = "1012",
    year = "2023"
}

@article{Guo:2023pqw,
    author = "Guo, Yuxun and Ji, Xiangdong and Liu, Yizhuang and Yang, Jinghong",
    title = "{Updated analysis of near-threshold heavy quarkonium production for probe of proton\textquoteright{}s gluonic gravitational form factors}",
    eprint = "2305.06992",
    archivePrefix = "arXiv",
    primaryClass = "hep-ph",
    doi = "10.1103/PhysRevD.108.034003",
    journal = "Phys. Rev. D",
    volume = "108",
    number = "3",
    pages = "034003",
    year = "2023"
}

@article{Won:2023ial,
    author = "Won, Ho-Yeon and Kim, Hyun-Chul and Kim, June-Young",
    title = "{Role of strange quarks in the D-term and cosmological constant term of the proton}",
    eprint = "2307.00740",
    archivePrefix = "arXiv",
    primaryClass = "hep-ph",
    reportNumber = "INHA-NTG-04/2023, JLAB-THY-23-3882",
    doi = "10.1103/PhysRevD.108.094018",
    journal = "Phys. Rev. D",
    volume = "108",
    number = "9",
    pages = "094018",
    year = "2023"
}

@article{Guo:2023qgu,
    author = "Guo, Yuxun and Ji, Xiangdong and Yuan, Feng",
    title = "{Proton\textquoteright{}s gluon GPDs at large skewness and gravitational form factors from near threshold heavy quarkonium photoproduction}",
    eprint = "2308.13006",
    archivePrefix = "arXiv",
    primaryClass = "hep-ph",
    doi = "10.1103/PhysRevD.109.014014",
    journal = "Phys. Rev. D",
    volume = "109",
    number = "1",
    pages = "014014",
    year = "2024"
}

@article{Czarnecki:2023yqd,
    author = "Czarnecki, Andrzej and Liu, Yizhuang and Reza, Syed Navid",
    title = "{Energy-momentum Tensor of a Hydrogen Atom: Stability, $D$-term, and the Lamb Shift}",
    eprint = "2309.10994",
    archivePrefix = "arXiv",
    primaryClass = "hep-ph",
    reportNumber = "Alberta Thy 7-23",
    doi = "10.5506/APhysPolBSupp.16.7-A19",
    journal = "Acta Phys. Polon. Supp.",
    volume = "16",
    number = "7",
    pages = "7--A19",
    year = "2023"
}

@article{Won:2023zmf,
    author = "Won, Ho-Yeon and Kim, Hyun-Chul and Kim, June-Young",
    title = "{Mechanical structure of the nucleon and the baryon octet: twist-2 case}",
    eprint = "2310.04670",
    archivePrefix = "arXiv",
    primaryClass = "hep-ph",
    reportNumber = "JLAB-THY-23-3938",
    doi = "10.1007/JHEP05(2024)173",
    journal = "JHEP",
    volume = "05",
    pages = "173",
    year = "2024"
}

@article{Hatta:2023fqc,
    author = "Hatta, Yoshitaka",
    title = "{Accessing the gravitational form factors of the nucleon and nuclei through a massive graviton}",
    eprint = "2311.14470",
    archivePrefix = "arXiv",
    primaryClass = "hep-ph",
    doi = "10.1103/PhysRevD.109.L051502",
    journal = "Phys. Rev. D",
    volume = "109",
    number = "5",
    pages = "L051502",
    year = "2024"
}

@article{Liu:2024rdm,
    author = "Liu, Wei-Yang and Shuryak, Edward and Zahed, Ismail",
    title = "{Glue in hadrons at medium resolution and the QCD instanton vacuum}",
    eprint = "2404.03047",
    archivePrefix = "arXiv",
    primaryClass = "hep-ph",
    doi = "10.1103/PhysRevD.110.054005",
    journal = "Phys. Rev. D",
    volume = "110",
    number = "5",
    pages = "054005",
    year = "2024"
}

@article{Yao:2024ixu,
    author = "Yao, Z. -Q. and Xu, Y. -Z. and Binosi, D. and Cui, Z. -F. and Ding, M. and Raya, K. and Roberts, C. D. and Rodr{\'\i}guez-Quintero, J. and Schmidt, S. M.",
    title = "{Nucleon gravitational form factors}",
    eprint = "2409.15547",
    archivePrefix = "arXiv",
    primaryClass = "hep-ph",
    reportNumber = "NJU-INP 091/24",
    doi = "10.1140/epja/s10050-025-01557-x",
    journal = "Eur. Phys. J. A",
    volume = "61",
    number = "5",
    pages = "92",
    year = "2025"
}

@article{Hatta:2025vhs,
    author = "Hatta, Yoshitaka and Klest, Henry T. and Passek-K., Kornelija and Schoenleber, Jakob",
    title = "{Deeply virtual $\phi$-meson production near threshold}",
    eprint = "2501.12343",
    archivePrefix = "arXiv",
    primaryClass = "hep-ph",
    doi = "10.1093/ptep/ptaf076",
    journal = {Progress of Theoretical and Experimental Physics},
    pages = {ptaf076},
    year = {2025},
    month = {05},
}

@article{Guo:2025jiz,
    author = "Guo, Yuxun and Yuan, Feng and Zhao, Wenbin",
    title = "{Bayesian Inferring Nucleon Gravitational Form Factors via Near-Threshold J/{\ensuremath{\psi}} Photoproduction}",
    eprint = "2501.10532",
    archivePrefix = "arXiv",
    primaryClass = "hep-ph",
    doi = "10.1103/3x7r-ythq",
    journal = "Phys. Rev. Lett.",
    volume = "135",
    number = "11",
    pages = "111902",
    year = "2025"
}

@article{Hatta:2025ryj,
    author = "Hatta, Yoshitaka and Schoenleber, Jakob",
    title = "{Sullivan Process near Threshold and the Pion Gravitational Form Factors}",
    eprint = "2502.12061",
    archivePrefix = "arXiv",
    primaryClass = "hep-ph",
    doi = "10.1103/y9fq-y84c",
    journal = "Phys. Rev. Lett.",
    volume = "134",
    number = "25",
    pages = "251901",
    year = "2025"
}

@article{Goharipour:2025lep,
    author = "Goharipour, Muhammad and Hashamipour, Hadi and Fatehi, H. and Irani, Fatemeh and Azizi, K. and Goloskokov, S. V.",
    collaboration = "MMGPDs",
    title = "{Mechanical properties of the nucleon from the generalized parton distributions}",
    eprint = "2501.16257",
    archivePrefix = "arXiv",
    primaryClass = "hep-ph",
    doi = "10.1103/jkhv-6949",
    journal = "Phys. Rev. D",
    volume = "112",
    number = "1",
    pages = "014016",
    year = "2025"
}

@article{Cebulla:2007ei,
    author = "Cebulla, C. and Goeke, K. and Ossmann, J. and Schweitzer, P.",
    title = "{The Nucleon form-factors of the energy momentum tensor in the Skyrme model}",
    eprint = "hep-ph/0703025",
    archivePrefix = "arXiv",
    reportNumber = "RUB-TP2-01-2007",
    doi = "10.1016/j.nuclphysa.2007.08.004",
    journal = "Nucl. Phys. A",
    volume = "794",
    pages = "87--114",
    year = "2007"
}

@article{Hudson:2017xug,
    author = "Hudson, Jonathan and Schweitzer, Peter",
    title = "{D term and the structure of pointlike and composed spin-0 particles}",
    eprint = "1712.05316",
    archivePrefix = "arXiv",
    primaryClass = "hep-ph",
    doi = "10.1103/PhysRevD.96.114013",
    journal = "Phys. Rev. D",
    volume = "96",
    number = "11",
    pages = "114013",
    year = "2017"
}

@article{Shanahan:2018pib,
    author = "Shanahan, P. E. and Detmold, W.",
    title = "{Gluon gravitational form factors of the nucleon and the pion from lattice QCD}",
    eprint = "1810.04626",
    archivePrefix = "arXiv",
    primaryClass = "hep-lat",
    reportNumber = "MIT-CTP/5069",
    doi = "10.1103/PhysRevD.99.014511",
    journal = "Phys. Rev. D",
    volume = "99",
    number = "1",
    pages = "014511",
    year = "2019"
}

@article{Krutov:2020ewr,
    author = "Krutov, A. F. and Troitsky, V. E.",
    title = "{Pion gravitational form factors in a relativistic theory of composite particles}",
    eprint = "2010.11640",
    archivePrefix = "arXiv",
    primaryClass = "hep-ph",
    doi = "10.1103/PhysRevD.103.014029",
    journal = "Phys. Rev. D",
    volume = "103",
    number = "1",
    pages = "014029",
    year = "2021"
}

@article{Hackett:2023nkr,
    author = "Hackett, Daniel C. and Oare, Patrick R. and Pefkou, Dimitra A. and Shanahan, Phiala E.",
    title = "{Gravitational form factors of the pion from lattice QCD}",
    eprint = "2307.11707",
    archivePrefix = "arXiv",
    primaryClass = "hep-lat",
    reportNumber = "MIT-CTP/5585",
    doi = "10.1103/PhysRevD.108.114504",
    journal = "Phys. Rev. D",
    volume = "108",
    number = "11",
    pages = "114504",
    year = "2023"
}

@article{Xu:2023izo,
    author = "Xu, Yin-Zhen and Ding, Minghui and Raya, Kh\'epani and Roberts, Craig D. and Rodr\'\i{}guez-Quintero, Jos\'e and Schmidt, Sebastian M.",
    title = "{Pion and kaon electromagnetic and gravitational form factors}",
    eprint = "2311.14832",
    archivePrefix = "arXiv",
    primaryClass = "hep-ph",
    reportNumber = "NJU-INP 081/23",
    doi = "10.1140/epjc/s10052-024-12518-x",
    journal = "Eur. Phys. J. C",
    volume = "84",
    number = "2",
    pages = "191",
    year = "2024"
}

@article{Fujii:2024rqd,
    author = "Fujii, Daisuke and Iwanaka, Akihiro and Tanaka, Mitsuru",
    title = "{Gravitational form factors of pion from top-down holographic QCD}",
    eprint = "2407.21113",
    archivePrefix = "arXiv",
    primaryClass = "hep-ph",
    doi = "10.1103/PhysRevD.110.L091501",
    journal = "Phys. Rev. D",
    volume = "110",
    number = "9",
    pages = "L091501",
    year = "2024"
}

@article{Liu:2024jno,
    author = "Liu, Wei-Yang and Shuryak, Edward and Weiss, Christian and Zahed, Ismail",
    title = "{Pion gravitational form factors in the QCD instanton vacuum. I}",
    eprint = "2405.14026",
    archivePrefix = "arXiv",
    primaryClass = "hep-ph",
    reportNumber = "JLAB-THY-24-4067",
    doi = "10.1103/PhysRevD.110.054021",
    journal = "Phys. Rev. D",
    volume = "110",
    number = "5",
    pages = "054021",
    year = "2024"
}

@article{Polyakov:2018zvc,
    author = "Polyakov, Maxim V. and Schweitzer, Peter",
    title = "{Forces inside hadrons: pressure, surface tension, mechanical radius, and all that}",
    eprint = "1805.06596",
    archivePrefix = "arXiv",
    primaryClass = "hep-ph",
    doi = "10.1142/S0217751X18300259",
    journal = "Int. J. Mod. Phys. A",
    volume = "33",
    number = "26",
    pages = "1830025",
    year = "2018"
}

@article{Burkert:2023wzr,
    author = "Burkert, V. D. and Elouadrhiri, L. and Girod, F. X. and Lorc\'e, C. and Schweitzer, P. and Shanahan, P. E.",
    title = "{Colloquium: Gravitational form factors of the proton}",
    eprint = "2303.08347",
    archivePrefix = "arXiv",
    primaryClass = "hep-ph",
    reportNumber = "JLAB-PHY-23-3774",
    doi = "10.1103/RevModPhys.95.041002",
    journal = "Rev. Mod. Phys.",
    volume = "95",
    number = "4",
    pages = "041002",
    year = "2023"
}

@article{Dehghan:2025eov,
    author = "Dehghan, Zeinab and Azizi, K.",
    title = "{Mechanical properties of the {\ensuremath{\Omega}}- baryon from gravitational form factors}",
    eprint = "2507.14840",
    archivePrefix = "arXiv",
    primaryClass = "hep-ph",
    doi = "10.1103/x6f5-cbcc",
    journal = "Phys. Rev. D",
    volume = "112",
    number = "5",
    pages = "054014",
    year = "2025"
}

@article{Coriano:2025lge,
    author = "Corian{\`o}, Claudio and Lionetti'', ''Stefano and Melle, Dario and Torcellini'', ''Leonardo and Tommasi, Riccardo",
    title = "{Gravitational Form Factors and the QCD Dilaton at Large Momentum Transfer}",
    eprint = "2504.20884",
    archivePrefix = "arXiv",
    primaryClass = "hep-ph",
    doi = "10.22323/1.490.0148",
    journal = "PoS",
    volume = "CORFU2024",
    pages = "148",
    year = "2025"
}

@misc{Sugimoto:2025btn,
    author = "Sugimoto, Shigeki and Tsukamoto, Taichi",
    title = "{Energy-Momentum Tensor and D-term of Baryons in Top-down Holographic QCD}",
    eprint = "2503.19492",
    archivePrefix = "arXiv",
    primaryClass = "hep-th",
    reportNumber = "KUNS-3037",
    doi = "10.1093/ptep/ptaf094",
    month = "3",
    year = "2025"
}

@article{Coriano:2024wrz,
    author = "Corian{\`o}, Claudio and Lionetti, Stefano and Melle, Dario and Tommasi, Riccardo",
    title = "{The gravitational form factor of the pion and proton and the conformal anomaly}",
    eprint = "2409.19586",
    archivePrefix = "arXiv",
    primaryClass = "hep-ph",
    doi = "10.1051/epjconf/202431400030",
    journal = "EPJ Web Conf.",
    volume = "314",
    pages = "00030",
    year = "2024"
}

@article{Cao:2025dkv,
    author = "Cao, Xiong-Hui and Guo, Feng-Kun and Li, Qu-Zhi and Wu, Bo-Wen and Yao, De-Liang",
    title = "{Gravitational form factors of pions, kaons and nucleons from dispersion relations}",
    eprint = "2507.05375",
    archivePrefix = "arXiv",
    primaryClass = "hep-ph",
    doi = "10.1140/epjs/s11734-025-02025-9",
    month = "7",
    journal = "Eur. Phys. J. ST",
    volume = "235",
    pages = "2121",
    year = "2026"
}

@article{Nair:2025sfr,
    author = "Nair, Sreeraj and Mondal, Chandan and Xu, Siqi and Zhao, Xingbo and Vary, James P.",
    collaboration = "BLFQ",
    title = "{Proton gravitational structure and mass decomposition on the light front}",
    eprint = "2506.07554",
    archivePrefix = "arXiv",
    primaryClass = "hep-ph",
    doi = "10.1103/79fs-5ghd",
    journal = "Phys. Rev. D",
    volume = "112",
    number = "11",
    pages = "114001",
    year = "2025"
}

@article{Liu:2025vfe,
    author = "Liu, Zhibo and Watanabe, Akira",
    title = "{Gravitational form factor of the kaon in holographic QCD}",
    eprint = "2503.18747",
    archivePrefix = "arXiv",
    primaryClass = "hep-ph",
    doi = "10.1103/1wbr-j3r4",
    journal = "Phys. Rev. D",
    volume = "112",
    number = "11",
    pages = "114032",
    year = "2025"
}

@article{Donoghue:1991qv,
    author = "Donoghue, John F. and Leutwyler, H.",
    title = "{Energy and momentum in chiral theories}",
    reportNumber = "UMHEP-321, NSF-ITP-91-119I",
    doi = "10.1007/BF01560453",
    journal = "Z. Phys. C",
    volume = "52",
    pages = "343--351",
    year = "1991"
}

@article{Liu:2023cse,
    author = "Liu, Keh-Fei",
    title = "{Hadrons, superconductor vortices, and cosmological constant}",
    eprint = "2302.11600",
    archivePrefix = "arXiv",
    primaryClass = "hep-ph",
    doi = "10.1016/j.physletb.2023.138418",
    journal = "Phys. Lett. B",
    volume = "849",
    pages = "138418",
    year = "2024"
}

@article{Cao:2024zlf,
    author = "Cao, Xiong-Hui and Guo, Feng-Kun and Li, Qu-Zhi and Yao, De-Liang",
    title = "{Dispersive determination of nucleon gravitational form factors}",
    eprint = "2411.13398",
    archivePrefix = "arXiv",
    primaryClass = "hep-ph",
    doi = "10.1038/s41467-025-62278-9",
    journal = "Nature Commun.",
    volume = "16",
    pages = "6979",
    year = "2025"
}

@article{Dehghan:2025ncw,
    author = "Dehghan, Zeinab and Almaksusi, F. and Azizi, K.",
    title = "{Mechanical properties of proton using flavor-decomposed gravitational form factors}",
    eprint = "2502.16689",
    archivePrefix = "arXiv",
    primaryClass = "hep-ph",
    doi = "10.1007/JHEP06(2025)025",
    journal = "JHEP",
    volume = "06",
    pages = "025",
    year = "2025"
}

@article{Goharipour:2025yxm,
    author = "Goharipour, Muhammad and Irani, Fatemeh and Amiri, M. H. and Fatehi, H. and Falahi, Behnam and Moradi, A. and Azizi, K.",
    collaboration = "MMGPDs",
    title = "{Can we determine the exact size of the nucleon?: A comprehensive study of different radii}",
    eprint = "2503.08847",
    archivePrefix = "arXiv",
    primaryClass = "hep-ph",
    doi = "10.1016/j.nuclphysb.2025.116962",
    journal = "Nucl. Phys. B",
    volume = "1017",
    pages = "116962",
    year = "2025"
}

@article{Tanaka:2025pny,
    author = "Tanaka, Mitsuru and Fujii, Daisuke and Kawaguchi, Mamiya",
    title = "{Gravitational form factors of the nucleon in the Skyrme model based on scale-invariant chiral perturbation theory}",
    eprint = "2507.21220",
    archivePrefix = "arXiv",
    primaryClass = "hep-ph",
    doi = "10.1103/gzj5-7bln",
    journal = "Phys. Rev. D",
    volume = "112",
    number = "5",
    pages = "054048",
    year = "2025"
}

@misc{Tanaka:2025hhv,
    author = "Tanaka, Kazuhiro",
    title = "{Proton mass decompositions in the NNLO QCD}",
    eprint = "2511.18310",
    archivePrefix = "arXiv",
    primaryClass = "hep-ph",
    month = "11",
    year = "2025"
}

@article{Voloshin:1982eb,
    author = "Voloshin, M. B. and Dolgov, A. D.",
    title = "{ON GRAVITATIONAL INTERACTION OF THE GOLDSTONE BOSONS}",
    journal = "Sov. J. Nucl. Phys.",
    volume = "35",
    pages = "120--121",
    year = "1982"
}

@article{Tanaka:2018wea,
    author = "Tanaka, Kazuhiro",
    title = "{Operator relations for gravitational form factors of a spin-0 hadron}",
    eprint = "1806.10591",
    archivePrefix = "arXiv",
    primaryClass = "hep-ph",
    reportNumber = "J-PARC-TH-0120",
    doi = "10.1103/PhysRevD.98.034009",
    journal = "Phys. Rev. D",
    volume = "98",
    number = "3",
    pages = "034009",
    year = "2018"
}

@article{Kumano:2017lhr,
    author = "Kumano, S. and Song, Qin-Tao and Teryaev, O. V.",
    title = "{Hadron tomography by generalized distribution amplitudes in pion-pair production process $\gamma^* \gamma \rightarrow \pi^0 \pi^0 $ and gravitational form factors for pion}",
    eprint = "1711.08088",
    archivePrefix = "arXiv",
    primaryClass = "hep-ph",
    reportNumber = "J-PARC-TH-0086, KEK-TH-1959, J-PARC-TH-0086 (Erratum: KEK-TH-2096, J-PARC-TH-0155)",
    doi = "10.1103/PhysRevD.97.014020",
    journal = "Phys. Rev. D",
    volume = "97",
    number = "1",
    pages = "014020",
    year = "2018"
}

@article{Wang:2024lrm,
    author = "Wang, Bigeng and He, Fangcheng and Wang, Gen and Draper, Terrence and Liang, Jian and Liu, Keh-Fei and Yang, Yi-Bo",
    collaboration = "{\ensuremath{\chi}}QCD",
    title = "{Trace anomaly form factors from lattice QCD}",
    eprint = "2401.05496",
    archivePrefix = "arXiv",
    primaryClass = "hep-lat",
    doi = "10.1103/PhysRevD.109.094504",
    journal = "Phys. Rev. D",
    volume = "109",
    number = "9",
    pages = "094504",
    year = "2024"
}

@article{Fujii:2025aip,
    author = "Fujii, Daisuke and Kawaguchi, Mamiya and Tanaka, Mitsuru",
    title = "{Dominance of gluonic scale anomaly in confining pressure inside nucleon and D-term}",
    eprint = "2503.09686",
    archivePrefix = "arXiv",
    primaryClass = "hep-ph",
    doi = "10.1016/j.physletb.2025.139559",
    journal = "Phys. Lett. B",
    volume = "866",
    pages = "139559",
    year = "2025"
}

@article{Broniowski:2025ctl,
    author = "Broniowski, Wojciech and Ruiz Arriola, Enrique",
    title = "{Gravitational form factors and mechanical properties of the nucleon in a meson dominance approach}",
    eprint = "2503.09297",
    archivePrefix = "arXiv",
    primaryClass = "hep-ph",
    doi = "10.1103/ylml-mrlh",
    journal = "Phys. Rev. D",
    volume = "112",
    number = "5",
    pages = "054028",
    year = "2025"
}

\end{document}